\documentclass[draft]{agujournal2019}
\usepackage{url} %this package should fix any errors with URLs in refs.
\usepackage{soul}
\usepackage[version=4]{mhchem}
\usepackage{xcolor}
\newcommand\nodata{ ~$\cdots$~ }
\makeatletter
\newcounter{reaction}
\renewcommand\thereaction{R\arabic{reaction}}
\newcommand\reactiontag%
{\refstepcounter{reaction}\tag{\thereaction}}
\newcommand\reaction@[2][]%
{\begin{equation}\ce{#2}%
\ifx\@empty#1\@empty\else\label{#1}\fi%
\reactiontag\end{equation}}
\newcommand\reaction@nonumber[1]%
{\begin{equation*}\ce{#1}\end{equation*}}
\newcommand\reaction%
{\@ifstar{\reaction@nonumber}{\reaction@}}
\makeatother

\draftfalse

\journalname{JGR: Planets}

\begin{document}

%% ------------------------------------------------------------------------ %%
%  Title
%
% (A title should be specific, informative, and brief. Use
% abbreviations only if they are defined in the abstract. Titles that
% start with general keywords then specific terms are optimized in
% searches)
%
%% ------------------------------------------------------------------------ %%

\title{Aerosols in Exoplanet Atmospheres}

%% ------------------------------------------------------------------------ %%
%
%  AUTHORS AND AFFILIATIONS
%
%% ------------------------------------------------------------------------ %%

% Authors are individuals who have significantly contributed to the
% research and preparation of the article. Group authors are allowed, if
% each author in the group is separately identified in an appendix.)

% List authors by first name or initial followed by last name and
% separated by commas. Use \affil{} to number affiliations, and
% \thanks{} for author notes.
% Additional author notes should be indicated with \thanks{} (for
% example, for current addresses).

% Example: \authors{A. B. Author\affil{1}\thanks{Current address, Antartica}, B. C. Author\affil{2,3}, and D. E.
% Author\affil{3,4}\thanks{Also funded by Monsanto.}}

\authors{Peter Gao\affil{1,2}, Hannah R. Wakeford\affil{3}, Sarah E. Moran\affil{4}, and Vivien Parmentier\affil{5}}

\affiliation{1}{Department of Astronomy and Astrophysics, University of California, Santa Cruz, CA 95064, USA}
\affiliation{2}{NHFP Sagan Fellow}
\affiliation{3}{School of Physics, University of Bristol, HH Wills Physics Laboratory, Tyndall Avenue, Bristol BS8 1TL, UK}
\affiliation{4}{Department of Earth and Planetary Sciences, Johns Hopkins University, Baltimore, MD 21218, USA}
\affiliation{5}{Department of Physics (Atmospheric, Oceanic and Planetary Physics), University of Oxford, Parks Rd, Oxford OX1 3PU, UK}

%% Corresponding Author:
% Corresponding author mailing address and e-mail address:

% (include name and email addresses of the corresponding author.  More
% than one corresponding author is allowed in this LaTeX file and for
% publication; but only one corresponding author is allowed in our
% editorial system.)

% Example: \correspondingauthor{First and Last Name}{email@address.edu}

\correspondingauthor{Peter Gao}{pgao8@ucsc.edu}

%% Keypoints, final entry on title page.

%  List up to three key points (at least one is required)
%  Key Points summarize the main points and conclusions of the article
%  Each must be 100 characters or less with no special characters or punctuation and must be complete sentences

% Example:
% \begin{keypoints}
% \item	List up to three key points (at least one is required)
% \item	Key Points summarize the main points and conclusions of the article
% \item	Each must be 100 characters or less with no special characters or punctuation and must be complete sentences
% \end{keypoints}

\begin{keypoints}
\item Aerosols are common in the atmospheres of exoplanets of all temperatures, masses, and compositions.  
\item Observations and models are painting a coherent picture of the nature of exoplanet aerosols.
\item Advances in laboratory work are essential for unveiling how exoplanet aerosols form and evolve.
\end{keypoints}

%% ------------------------------------------------------------------------ %%
%
%  ABSTRACT and PLAIN LANGUAGE SUMMARY
%
% A good Abstract will begin with a short description of the problem
% being addressed, briefly describe the new data or analyses, then
% briefly states the main conclusion(s) and how they are supported and
% uncertainties.

% The Plain Language Summary should be written for a broad audience,
% including journalists and the science-interested public, that will not have 
% a background in your field.
%
% A Plain Language Summary is required in GRL, JGR: Planets, JGR: Biogeosciences,
% JGR: Oceans, G-Cubed, Reviews of Geophysics, and JAMES.
% see http://sharingscience.agu.org/creating-plain-language-summary/)
%
%% ------------------------------------------------------------------------ %%

%% \begin{abstract} starts the second page

\begin{abstract}

Observations of exoplanet atmospheres have shown that aerosols, like in the Solar System, are common across a variety of temperatures and planet types. The formation and distribution of these aerosols are inextricably intertwined with the composition and thermal structure of the atmosphere. At the same time, these aerosols also interfere with our probes of atmospheric composition and thermal structure, and thus a better understanding of aerosols lead to a better understanding of exoplanet atmospheres as a whole. Here we review the current state of knowledge of exoplanet aerosols as determined from observations, modeling, and laboratory experiments. Measurements of the transmission spectra, dayside emission, and phase curves of transiting exoplanets, as well as the emission spectrum and light curves of directly imaged exoplanets and brown dwarfs have shown that aerosols are distributed inhomogeneously in exoplanet atmospheres, with aerosol distributions varying significantly with planet equilibrium temperature and gravity. Parameterized and microphysical models predict that these aerosols are likely composed of oxidized minerals like silicates for the hottest exoplanets, while at lower temperatures the dominant aerosols may be composed of alkali salts and sulfides. Particles originating from photochemical processes are also likely at low temperatures, though their formation process is highly complex, as revealed by laboratory work. In the years to come, new ground- and space-based observatories will have the capability to assess the composition of exoplanet aerosols, while new modeling and laboratory efforts will improve upon our picture of aerosol formation and dynamics.  

\end{abstract}

\section*{Plain Language Summary}
For nearly two decades we have had the opportunity to probe the atmospheres of planets orbiting other stars (``exoplanets''). These efforts have revealed the existence of clouds and hazes in these atmospheres, which prevent us from learning more about exoplanet atmospheres as a whole by blocking us from probing parts of the atmosphere below the cloud and haze layers. Here we summarize our current understanding of these structures. Using data from telescopes on the ground and in space, we have found that exoplanet clouds are patchy and are distributed mostly according to the temperature of the local atmosphere. Using computer simulations we have surmised that these clouds are likely made of materials that make up rocks on Earth, as the exoplanets we have probed thus far orbit their stars closely, resulting in very high temperatures in their atmospheres. At lower temperatures, but still several hundred degrees above room temperature, hazes composed of organic material are possible. These hazes are likely formed from complex chemical reactions, which are the current focus of laboratory experiments. Future efforts in data collection, computer simulations, and lab work will lead to a better understanding of exoplanet clouds and hazes. 

\section{Introduction}

Aerosols are fundamental components of planetary atmospheres. Every perennial atmosphere in the Solar System, including that of Pluto, Saturn's moon Titan, Neptune's moon Triton, and every planet except Mercury possess some form of such small suspended particulates. The composition of these aerosols is extremely diverse, including sulfuric acid on Venus \cite{hansen1974}, water on Earth, water, mineral dust, and carbon dioxide on Mars \cite{montmessin2007}, ammonia on Jupiter and Saturn \cite{brooke1998,baines2009}, and complex organics and condensed hydrocarbons and nitriles on Uranus, Neptune, Titan, Triton, and Pluto {\cite{sromovsky2011,romani1988,sagan1992,brown2002,rages1992,gladstone2016,wong2017,lavvas2020,ohno2021}}. In addition, these aerosols are inexorably tied to the chemistry, dynamics, and radiative environments of their host atmospheres. Sulfuric acid clouds on Venus form a vital branch of its sulfur chemical cycle and provide the planet its high albedo \cite{mills2007} while water clouds on Earth and dust on Mars are strong controls of their surface climates \cite{rosenfeld2014,martinez2017}. In the outer solar system, moist convection on the giant planets with the condensation of ammonia, water, and methane sculpts their global atmospheric dynamics and trace gas distributions \cite{lunine1993,hueso2006,li2015,bolton2017}; organic hazes on Titan and Pluto are the products of complex chemical networks and are major contributors to heating and cooling rates in their atmospheres \cite{mckay1989,zhang2017}; and latent heat release from nitrogen condensation on Triton could control its atmospheric thermal structure \cite{rages1992}. Understanding the formation and impact of aerosols on solar system objects have thus been vital for understanding their atmospheres as a whole. The same applies to exoplanets. 

Aerosols were anticipated to exist in exoplanet atmospheres not long after the discovery of the first exoplanet orbiting a sun-like star \cite{guillot1996,saumon1996}. In the few years that followed, several works \cite{burrows1997,seager1998,marley1999exo,seager2000,seager2000curve,sudarsky2000,hubbard2001,barman2001,sudarsky2003,baraffe2003} considered the formation of mineral and metal aerosols, e.g. silicates and iron, in one-dimensional (vertical), globally averaged models, as inspired by earlier and contemporary studies into equilibrium condensation and cloud formation processes in brown dwarf atmospheres \cite{lunine1989,marley1996,allard1997,tsuji1999,lodders1999,chabrier2000,burrows2000,allard2001,ackerman2001,helling2001,marley2002,lodders2002tiv,burrows2002,tsuji2002,cooper2003,woitke2003}, which possess similar temperatures ($\geq$1000 K) to those of the first exoplanets found. These early studies clearly demonstrated the importance of high temperature aerosol formation on the composition of the atmosphere and the planets' albedo, optical phase curve, polarization, and transmission, reflected light, and emission spectra. 

Observational evidence for aerosols in exoplanet atmospheres arrived with the measurement of the first exoplanet transmission spectrum - light from the host star filtered through the atmosphere of the planet during transit captured at a range of wavelengths. With this, \citeA{charbonneau2002} showed that absorption by atomic sodium in the atmosphere of the hot Jupiter HD\,209458b was less than predicted in clear atmosphere models, suggesting the existence of a layer of aerosols at high altitudes that obscured the wings of the sodium doublet \cite{fortney2003}. Subsequent transmission spectroscopy of another hot Jupiter, HD\,189733b at optical \cite{pont2008} and infrared wavelengths \cite{tinetti2007} showed a non-detection of alkali metal absorption lines and a significant offset in the transit radius between the two wavelength ranges; this again suggested the existence of high altitude aerosols that became optically thin at wavelengths $\geq$1\,$\mu$m \cite{lecavelierdesetangs2008}. Aerosols were also inferred in the atmospheres of some of the first directly imaged young giant exoplanets due to their red infrared colors \cite{marois2008}. These initial observations were the first hints that aerosols were just as ubiquitous in exoplanet atmospheres as in the atmospheres of our solar system worlds. 

In this review, we will summarize our current understanding of exoplanet aerosols, focusing primarily on advancements in knowledge made in the 2010s. These advancements include (1) the proliferation of exoplanet transmission spectroscopy, reflected light and emission photometry, and observations of exoplanet phase curves, which can all be used to probe exoplanet aerosols, (2) greater synergy between exoplanet and brown dwarf science with a focus on photometry and spectroscopy of directly imaged exoplanets, (3) development of more rigorous aerosol models in 1D and the extension to 3D, and (4) the application of laboratory experiments to investigate exoplanet aerosol formation and corresponding impact on observations. We refer the reader to \citeA{helling2008a,marley2013} for comprehensive reviews of pre-2010 studies. 

We begin by defining specific types of aerosols based on their formation processes and describing their possible compositions in ${\S}$\ref{sec:def} to facilitate our discussions. In ${\S}$\ref{sec:obs} we will provide an overview of the insights into exoplanet aerosols we have gained through observing exoplanet atmospheres in transmission, emission, and reflection. We will then focus on the usage of and predictions made by various aerosol models in ${\S}$\ref{sec:theo} and the complexities in aerosol formation revealed by laboratory work in ${\S}$\ref{sec:lab}. Finally, in ${\S}$\ref{sec:pic} we describe an emerging picture of how aerosol properties vary among different exoplanets and discuss how our understanding of exoplanet aerosols can progress in the years to come.

\section{Aerosol Definitions, Composition, and Provenance}\label{sec:def}

\begin{figure}[hbt!]
\centering
\includegraphics[width=1.0 \textwidth]{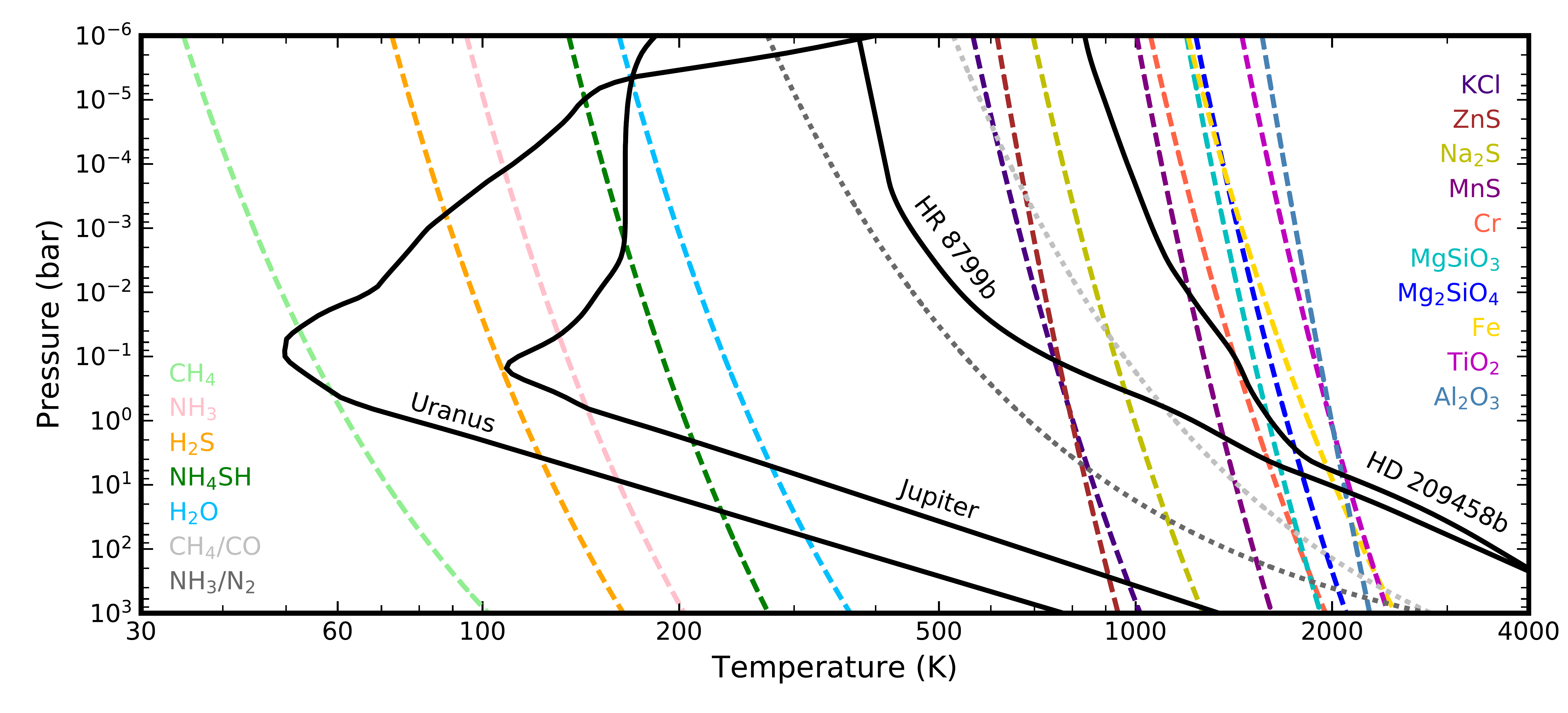}
\caption{Condensation temperatures of various cloud species as a function of atmospheric pressure, assuming solar metallicity, compared to temperature-pressure (TP) profiles of several objects. {Condensation of a given species can occur when the planet TP profile becomes lower than its condensation temperature profile.} TP profiles for Jupiter and Uranus are taken from \citeA{moses2017} while those of HR 8799b and HD 209458b are generated by a thermal structure model \cite{saumon2008} assuming appropriate planetary parameters. The condensation curve for CH$_4$ is computed by combining the CH$_4$ saturation vapor pressure \cite{lodders1998} with its mixing ratio in a solar metallicity gas \cite{lodders2010}, assuming that all carbon is in the form of CH$_4$. The condensation curves for NH$_3$, NH$_4$SH, and H$_2$O are taken from \citeA{lodders2002a}; that of H$_2$S is from \citeA{visscher2006}; those of KCl, ZnS, Na$_2$S, MnS, and Cr are from \citeA{morley2012}; those of MgSiO$_3$, Mg$_2$SiO$_4$, and Fe are from \citeA{visscher2010}; that of TiO$_2$ is from \citeA{helling2001}; and that of Al$_2$O$_3$ is from \citeA{wakeford2017refractory}. The CH$_4$/CO and NH$_3$/CO transition curves are from \citeA{lodders2002a}. }
\label{fig:cloudcomp}
\end{figure}

%(Right) Chemical species mixing ratio profiles from photochemical simulations by \citeA{kawashima2019b} for their 100 $\times$ solar metallicity case (see their Figure 10c).

A number of terms have been used to refer to aerosols in planetary and exoplanet atmospheres in the literature, including clouds, hazes, and dust. For clarity, we will assign to them specific definitions based on their provenance in this review, inspired by \citeA{horst2016}. Where provenance is unclear, we will use the catch-all term, ``aerosols.'' 

\paragraph*{Dust:}
We define dust as particles lifted into the atmosphere from a planetary surface, such as sand and sea salt on Earth, fine regolith particles on Mars, and organic dune particles and ices on Titan and Pluto. 

\paragraph*{Clouds:}
We define clouds as collections of particles forming in the atmosphere under thermochemical equilibrium. This definition includes both first order phase changes, such as 

%\begin{equation}
%\label{eq:water}
\reaction{H2O(g)  <--> H2O(s,l)}
%\reaction{CO2 + C2}
%\reactionnonumber{CO2 + C}

%{\rm H_2O_{(g)}} \longleftrightarrow {\rm H_2O_{(s,l)}} 
%\end{equation}

\noindent as well as thermochemical reactions like 

\reaction{2Mg(g) + 3H2O(g) + SiO(g) <--> Mg2SiO4(s,l) + 3H2(g)}

%\begin{equation}
%\label{eq:forsterite}
%{\rm 2Mg_{(g)}} + {\rm 3H_2O_{(g)}} + {\rm SiO_{(g)}} \longleftrightarrow {\rm Mg_2SiO_{4(s,l)}} + {\rm 3H_{2(g)}}
%\end{equation}

%For this audience, it is fine to say that this is "thermochemical equilibrium" that arises from a "minimization of Gibbs free energy", given the local temperature, pressure, and elemental abundances. E.g., define TEQ as the state at which the GFE is minimized, as you use the term "thermochemical equilibrium" in the subsequent paragraph.

%In other words, clouds form when it is energetically favorable for gas molecules to spontaneously cluster themselves into or react with each other to form solids and/or liquids given the local chemical composition, temperature, and pressure. 

\noindent Thermochemical equilibrium arises from the minimization of Gibbs free energy given the local temperature, pressure, and elemental abundances. Because of this, cloud formation is locally reversible, such that the loss of clouds through evaporation or chemical decomposition is in balance with condensation and synthesis. In the Solar System, clouds tend to form via condensation, a first order phase change, such as those of water, carbon dioxide, ammonia, methane, and nitrogen. Meanwhile, ammonium hydrosulfide (NH$_4$SH) clouds, for which we have indirect evidence for in the atmospheres of the giant planets, form through chemical reactions between gaseous ammonia and hydrogen sulfide \cite<e.g.,>{lewis1969,carlson1988,depater2014,wong2015fresh,bjoraker2018}. 

Thermochemical equilibrium models have predicted a myriad of cloud compositions in exoplanet atmospheres under the assumption that the atmospheric gas composition is 1 to only several times more enriched in metals than a solar composition gas. {``Metals'' in this case refers to all elements heavier than hydrogen and helium} \cite<Figure \ref{fig:cloudcomp}; see e.g.>{burrows1999,lodders1999,lodders2002tiv,visscher2006,visscher2010,wakeford2017refractory}. Some of the proposed clouds form via phase changes, e.g. iron, chromium, potassium chloride, while others form via chemical reactions, e.g. forsterite, enstatite, corundum, and various sulfides. These compositions can vary significantly at higher metallicities and/or different carbon-to-oxygen ratios, such as the formation of clouds of graphite and carbides at high C/O \cite{moses2013c,mbarek2016,helling2017}. Additional cloud compositions can arise from the condensation of gases produced from photochemistry, such as sulfuric acid on Venus, hydrocarbons and nitriles on Titan and Pluto, and elemental sulfur in reducing atmospheres \cite<e.g.>{hu2013,zahnle2016}. 

\paragraph*{Hazes:}
Clouds derived from gases originating from photochemisty are distinct from hazes, which we define as collections of particles formed \textit{directly} from energy input via photochemistry and energetic particle bombardment. Hazes form through these processes via the breakdown of simple molecules like methane, nitrogen, carbon monoxide, hydrogen sulfide, etc. at low pressures to create radicals and ions, which then react to build more complex species, eventually forming small particles through successive reactions \cite{trainer2013,lavvas2013,yoon2014}. As a result, the exact compositions of hazes are highly uncertain, though their elemental make-up reflects the major gases in the atmosphere. Unlike clouds, haze formation is locally irreversible, tending towards complexity due to the external input of energy. Examples of hazes in the Solar System include those of Titan, Pluto, and the giant planets, as well as smog in Earth cities. Hazes in exoplanet atmospheres, particularly at high temperatures, could be more complex since they can incorporate elements that would otherwise be hidden in deep clouds on cooler worlds, including sodium, potassium, chlorine, magnesium, and iron. 

%The composition of exoplanet clouds have been largely estimated from equilibrium chemistry. By minimizing the Gibbs free energy of a parcel of gas 

%Clouds formed from the condensation of photochemical products are distinct from hazes, which we define as collections of particles formed \textit{directly} from energy input via photochemistry and energetic particle bombardment.

%It is important to note that aerosols formed from the condensation of gases that are themselves generated through photochemistry/energetic particle bombardment, e.g. sulfuric acid on Venus, ethane on Titan, and elemental sulfur in reducing atmospheres \cite<e.g.>{hu2013,zahnle2016} are not hazes, but clouds; hazes refer only to particles generated directly from energetic reactions.

%Some aerosols can fit both our cloud and haze definitions. For example, gaseous ethane on Titan and sulfuric acid vapor on Venus both condense to form aerosols, but they are also both products of photochemistry, sourced from methane and sulfur dioxide, respectively. Because the formation of these aerosols ultimately depends on whether condensation is energetically favored, we refer to them as clouds, and reserve ``hazes'' for aerosols where condensation is not a major formation pathway.

Our definitions of clouds, hazes, and dust are different from some of the previous and current usages of these terms in the exoplanet literature. For example, dust has been used to refer to high temperature condensates like iron and silicates as they likely condense to solid particles \cite{seager1998,ebel2006,pont2013}, but as they form via condensation and thermochemical reactions of atmospheric gasses these would be referred to as clouds under our definition regardless of their phase. By specifying dust as originating from a planetary surface due to mechanical processes like weathering, we relegate them to only thin atmospheres, such as those of rocky exoplanets. In addition, when interpreting observations, clouds and hazes have been used to refer to particles of different sizes and/or structures of different optical depths, independent of their formation processes. In transmission spectroscopy in the optical and near-infrared, large particles that act as gray absorbers/scatterers are often labeled as clouds, while small particles that preferentially scatter short wavelength visible light are labeled as hazes \cite{pont2008,sing2016,barstow2017,goyal2018}. Hazes have also been used to describe low optical depth aerosol layers above more optically thick ``cloud'' layers \cite{yang2015}. These uses are convenient for differentiating between the effects different aerosols have on observations when aerosol provenance is unknown. However, as we probe exoplanet atmospheres with more advanced instruments and more sophisticated techniques \cite{wakeford2015,kempton2017,powell2019}, aerosol definitions that are more connected to their formation mechanisms and underlying atmospheric processes will become increasingly informative. 

%ADD IN ROCKY PLANET CONDENSATES N2, CO2, H2SO4? 

\section{Insights from Observations}\label{sec:obs}

\begin{figure}[hbt!]
\centering
\includegraphics[width=1.0 \textwidth]{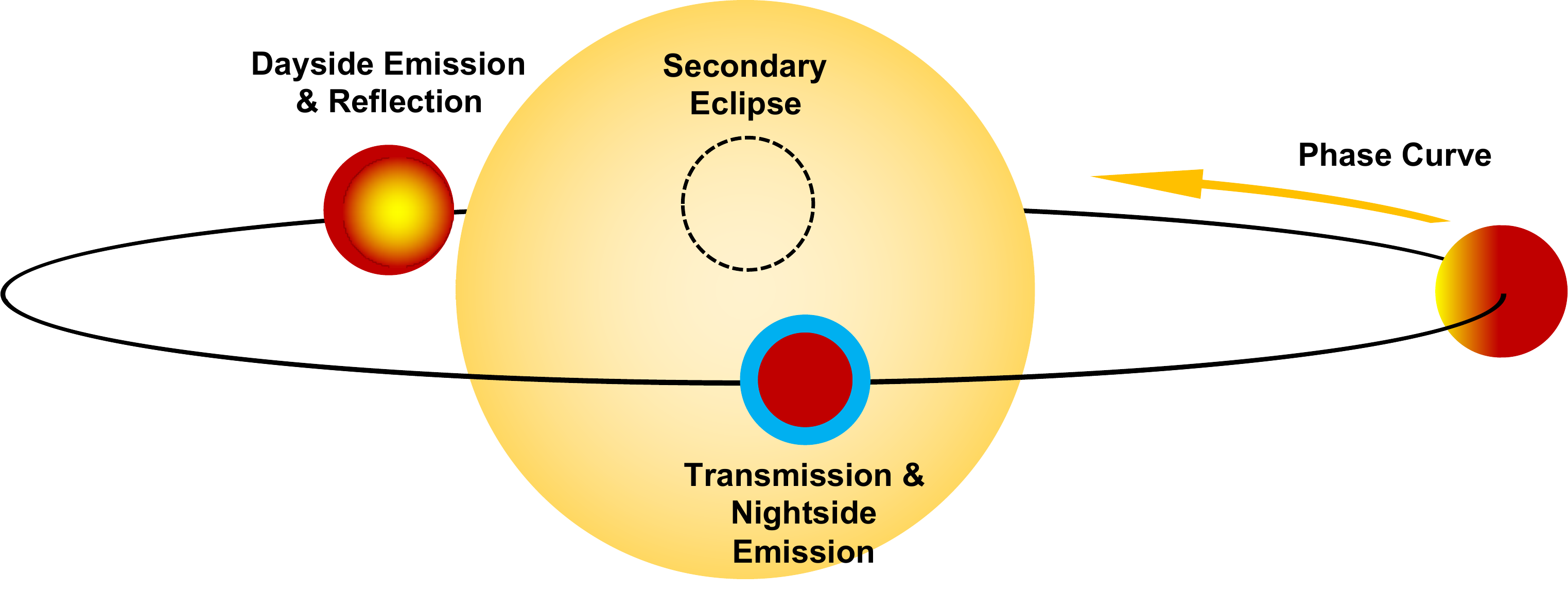}
\caption{The geometry of a transiting exoplanet as seen from Earth. When the exoplanet passes in front of its star with respect to us, we can measure the transmission spectrum of the limb of its atmosphere thanks to light from the host star filtering through the atmosphere on its way to us, as well as thermal emission from the nightside. When the exoplanet passes behind its star during secondary eclipse, its dayside is blocked; comparison between the total brightness of the exoplanet-star system before/after and during the secondary eclipse then allows for the measurement of the dayside flux, which is a combination of reflected star light and thermal emission. During the rest of the exoplanet's orbit, reflected star light and thermal emission as a function of longitude can be measured by observing the exoplanet's phase curve. The figure is not to scale.}
\label{fig:exoconfig}
\end{figure}

Aerosols impact every method of exoplanet atmosphere characterization (Figure \ref{fig:exoconfig}). Aerosol opacity controls the pressure levels probed in transmission through the atmosphere, emission from the atmosphere, and reflection by the atmosphere, suppressing the spectral signatures of molecular species at higher pressures. Heating and cooling by aerosols change the atmospheric temperature profile, regulating the planet's emitted flux. The reflectivity of aerosols, controlled by their optical constants, size, and shape, determines the albedo of a planet and thus the reflected light spectrum. Aerosol scattering and absorption also generate their own features in exoplanet spectra. In this section, we describe what observations of exoplanet atmospheres have revealed about exoplanet aerosols. 

\subsection{Aerosols in Transmission}

\begin{figure}[hbt!]
\centering
\includegraphics[width=0.9 \textwidth]{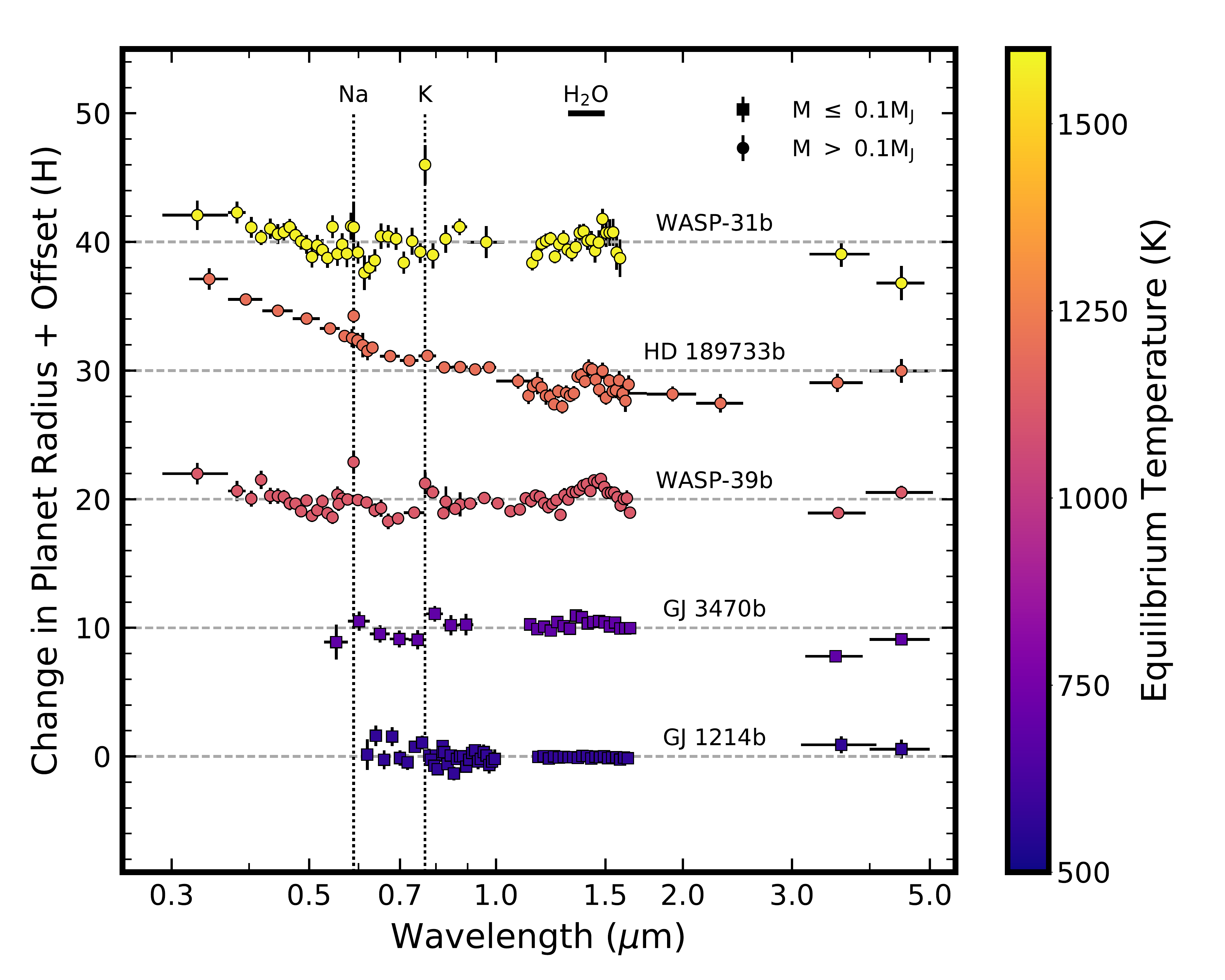}
\caption{Transmission spectra of several exoplanets showing the various impacts of aerosols. Planets with masses $>$0.1 Jupiter masses are shown in circles, while lower mass planets are shown in squares. The colors of the datasets represent the equilibrium temperatures of the corresponding planet. Observations are taken from \citeA{bean2011,desert2011,kreidberg2014,sing2015,sing2016,benneke2019,wakeford2018}. The transmission spectra are offset for clarity, normalised to the mean transit depth, and shown in planetary scale heights calculated using parameters listed on exo.MAST (e.g. gravity and equilibrium temperature), assuming solar metallicity and atmospheres dominated by H/He, i.e. an atmospheric mean molecular weight of 2.3 g mol$^{-1}$. {We refer the reader to https://stellarplanet.org/science/exoplanet-transmission-spectra/ for an up to date database of published exoplanet transmission spectra.}}
\label{fig:tspec}
\end{figure}

Transmission spectroscopy probes the day-to-night terminator of planetary atmospheres. Although it has become the most prolific method by which we probe exoplanet atmospheres \cite{kreidberg2018review}, it also leads to the most complex results to interpret, due to the large thermal and wind gradients across the atmospheric limb. The slant optical path through the atmosphere tangential to the target planet afforded by the observational geometry allows for probes of minute abundances of both molecular species ($\sim$1-100 ppm) and aerosol particles \cite{fortney2005cond}. Transmission spectra of a variety of exoplanets ranging from hot Jupiters to cool rocky worlds have been observed from the near-UV to the mid-infrared by ground- and space-based instruments \cite<e.g.>{charbonneau2002,bean2010,desert2011,wood2011,gibson2012,deming2013,swain2013,crossfield2013,knutson2014a,kreidberg2014,sing2015,dewit2016,wakeford2017wasp101b,kreidberg2018,bruno2018,kilpatrick2018,benneke2019k218b,chachan2019,libbyroberts2020,thao2020,wakeford2020}, all of which have been shown to be impacted by aerosols. The effect of aerosols span a continuum, from increased scattering slopes and reduced molecular features to completely featureless spectra (Figure \ref{fig:tspec}).

\begin{figure}[hbt!]
\centering
\includegraphics[width=1.0 \textwidth]{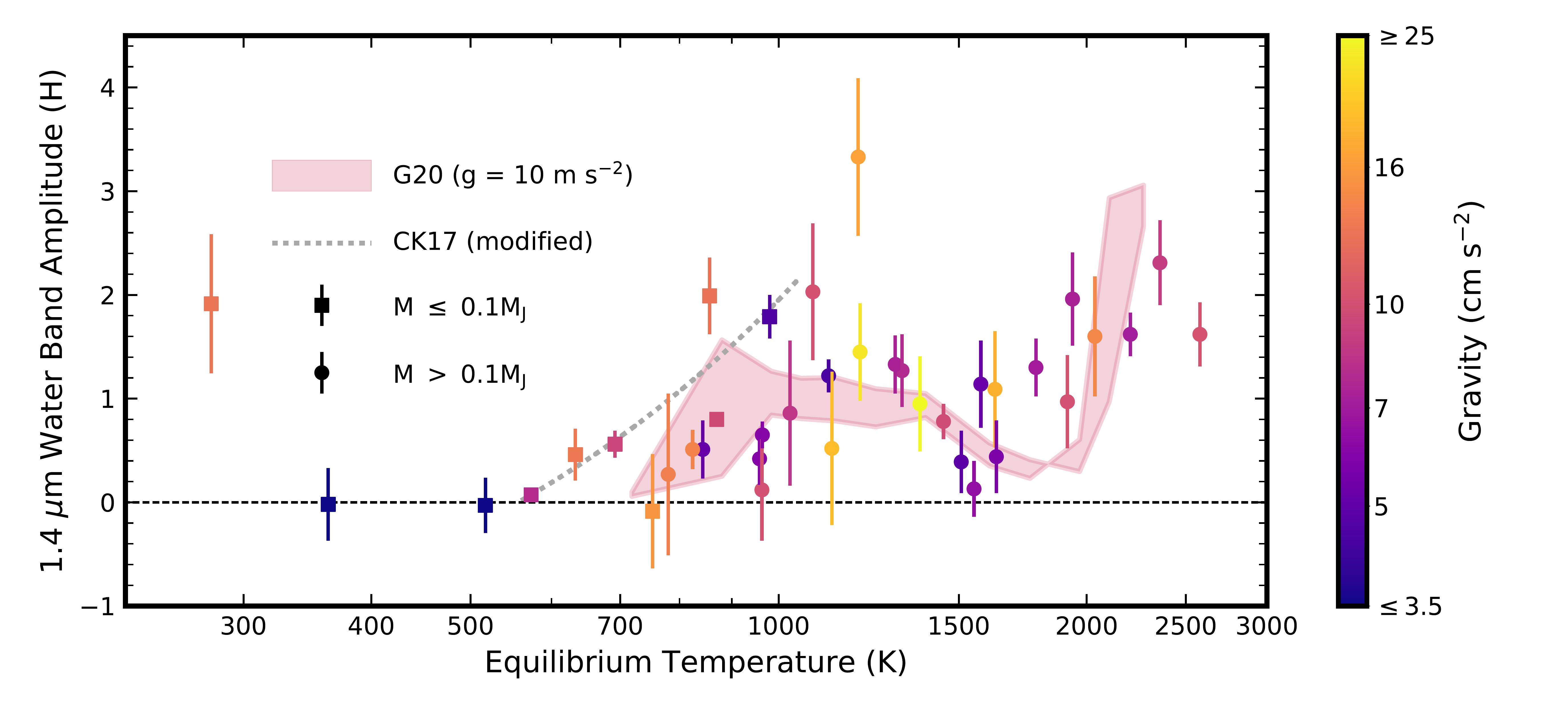}
\caption{Amplitude of the 1.4 $\mu$m water feature in transmission spectra of exoplanets in units of atmospheric scale height $H$ (defined as in Figure \ref{fig:tspec}) as a function of gravity and equilibrium temperature. Planets with masses $>$0.1 Jupiter masses are shown as circles, with the water feature amplitude values taken from \citeA{fu2017}, while lower mass planets are shown as squares. Water feature amplitudes for lower mass planets are taken from \citeA{crossfield2017,libbyroberts2020,kreidberg2020}. The water feature amplitude for K2-18b in units of $H$, 1.915$\pm$0.67, is calculated from the transmission spectrum presented in \citeA{benneke2019k218b} following the method of \citeA{stevenson2016}. The predicted water feature amplitude from \citeA{gao2020} for objects with gravity of 10 m s$^{-2}$ and atmospheric metallicity between 1 and 10 $\times$ solar is shown in pink. The best fit linear trend to the \citeA{crossfield2017} data, modified from the original publication to take into account the slightly different definition of equilibrium temperature, is shown in the dotted line, and has the functional form of A = 0.0044T - 2.45, where A is the water feature amplitude in units of $H$ and T is the temperature in K. }
\label{fig:h2oamp}
\end{figure}

The relatively large number ($\sim$50) of published transmission spectra have allowed for population studies of exoplanet atmospheres. Several studies have focused on the amplitude of the 1.4 $\mu$m water absorption feature above its adjacent low opacity regions at $\sim$1.2 and 1.6\,$\mu$m (equivalent to the J and H bands), modulated by the atmospheric scale height, as a measure of the vertical extent of the aerosols in the atmospheres of these planets \cite<Figure \ref{fig:h2oamp};>{sing2016,stevenson2016,iyer2016,fu2017,crossfield2017,tsiaras2018,wakeford2019}. {The scale height is typically computed using an atmospheric mean molecular weight corresponding to solar metallicity (2.3 g mol$^{-1}$) and a temperature equal to the planets' equilibrium temperature ($T_{eq}$). Higher metallicities are certainly possible, particularly for the lower mass planets, in which case assumptions of a solar metallicity scale height would underestimate the number of scale heights spanned by the observed water feature. The use of $T_{eq}$ for defining the scale height is also approximate, since the temperature at the altitudes probed by transits could be higher or lower.} These studies have found that aerosols diminish the water feature amplitude with respect to a clear atmosphere by 50-70\% on average. Some studies \cite{stevenson2016,fu2017} have claimed that there may be a trend in water feature amplitude with $T_{eq}$, from $\sim$600 to $\sim$2500 K, where hotter planets tend to have larger water feature amplitudes, suggesting deeper and/or more optically thin aerosol layers, while planet gravity appears to not significantly affect the feature. \citeA{crossfield2017,libbyroberts2020} showed that a similar trend may exist for Neptune-mass and sub-Neptune planets with $T_{eq}$ $<$ 1000 K, though higher atmospheric metallicity (i.e. higher mean molecular weight atmospheres) could also be responsible. However, this is in contrast with the recent discovery of a high amplitude water feature for K2-18b, a sub-Neptune with Earth-like temperatures \cite{benneke2019k218b,tsiaras2019}. Moreover, \citeA{fisher2018} found no significant trend between temperature and aerosol opacity in their combined analysis of giant and lower mass exoplanet near-infrared transmission spectra.

%Recent observations by \citeA{benneke2019k218b,tsiaras2019,libbyroberts2020} extended the exoplanet sample to Earth-like temperatures and suggest that the temperature trend may reverse itself below $\sim$300 K, though more observations of cool exoplanets are needed. 

At optical wavelengths, \citeA{heng2016} measured the amplitude of the atomic sodium and potassium absorption peaks and also claims a possible ``cloudiness'' trend with $T_{eq}$, with higher temperature planets being clearer. \citeA{sing2016} considered the relative increase in transit depths in the optical versus the mid-infrared as a measure of extinction by small particles, and found that aerosols are the primary factor that shape transmission spectra rather than variable water abundance. Also, several studies {\cite{pinhas2019,welbanks2019,may2020,alderson2020,chen2021}} have measured optical spectral slopes steeper than that of Rayleigh scattering. This would require an opacity source that varies with altitude, such as highly scattering aerosols with variable particle size and/or vertical distribution \cite{lecavelierdesetangs2008,sing2011,wakeford2015}.

In addition to focusing on specific wavelength regions and spectral features, a number of studies have performed uniform, homogeneous retrieval analyses on a large number of planets' complete transmission spectra from optical to mid-infrared wavelengths. {Retrievals are data-model parameter estimation procedures commonly used to infer the state properties (abundances, temperatures, cloud properties) of an atmosphere given a spectrum.} Both \citeA{barstow2017} and \citeA{pinhas2019} performed a retrieval on the 10 planets presented in \citeA{sing2016}. Although they both propose that non-monotonic trends with temperature exist in aerosol coverage at the limb of hot Jupiters, their results are incompatible. This highlights the sensitivity of retrieval studies to the details of the cloud parametrization \cite{barstow2020cloud} and the many degeneracies present between aerosol physical parameters such as altitude range, latitudinal coverage, particle size distribution, etc. (see ${\S}$\ref{sec:mod}).

%applied the NEMESIS optimal estimation framework to the hot Jupiter transmission spectra presented by \citeA{sing2016} and found that the trend in aerosol distribution with temperature may not be monotonic, with deeper aerosol layers (clearer atmospheres) between 1300 and 1700 K and more high-altitude aerosols (``cloudier'' atmospheres) at both cooler and warmer temperatures. This contrasts with the Bayesian retrieval conducted by \citeA{pinhas2019} on the same planets, which suggested the opposite trend, with the existence of high altitude aerosols on planets with temperatures between 1400 and 1600 K, and deeper aerosol layers for planets at lower and higher temperatures. This difference is likely due to the different aerosol parameterizations detailed by \citeA{Barstow2020cloud} (see ${\S}$\ref{sec:mod}). 

\subsection{Aerosols in Emission and Reflection}

%Sounds like we could to an explicit brown dwarfs/directly imaged planets / transiting planets subsections here. There's a whole paragraph on brown dwarf that is kind of unexpected. See my comment Vivian I agree here. I suggest a split based on sub-stellar vs planet

Thermal emission has been detected from two distinct populations of exoplanets: directly imaged young giant exoplanets in wide orbits about their host stars and transiting worlds on tight orbits ranging from hot Jupiters to rocky planets. {Emission from a handful of non-transiting exoplanets have also been detected \cite{harrington2006,crossfield2010,brogi2012,lockwood2014,piskorz2016,piskorz2017,birkby2017,webb2020}, but these observations have not yet been used to infer aerosol properties.} The nadir geometry of emission observations allow us to probe deeper into the atmosphere than transmission, with the emitted flux being a sensitive function of atmospheric thermal structure in addition to chemical composition and aerosol distribution. Thermal emission observations capture the average of the properties of a full hemisphere (often the dayside for transiting exoplanets), which makes them less sensitive to small variations in cloud properties and cloud coverage than transmission spectra. Thermal emission observed over a significant fraction of the rotation period of the object also gives information on longitudinal heterogeneity in the atmosphere; for transiting exoplanets, this is accomplished by observing the orbital phase curve over a significant fraction or all of the orbital period, as they are tidally locked to their stars, i.e. their rotation and orbital periods are the same. 

\begin{figure}[hbt!]
\centering
\includegraphics[width=0.8 \textwidth]{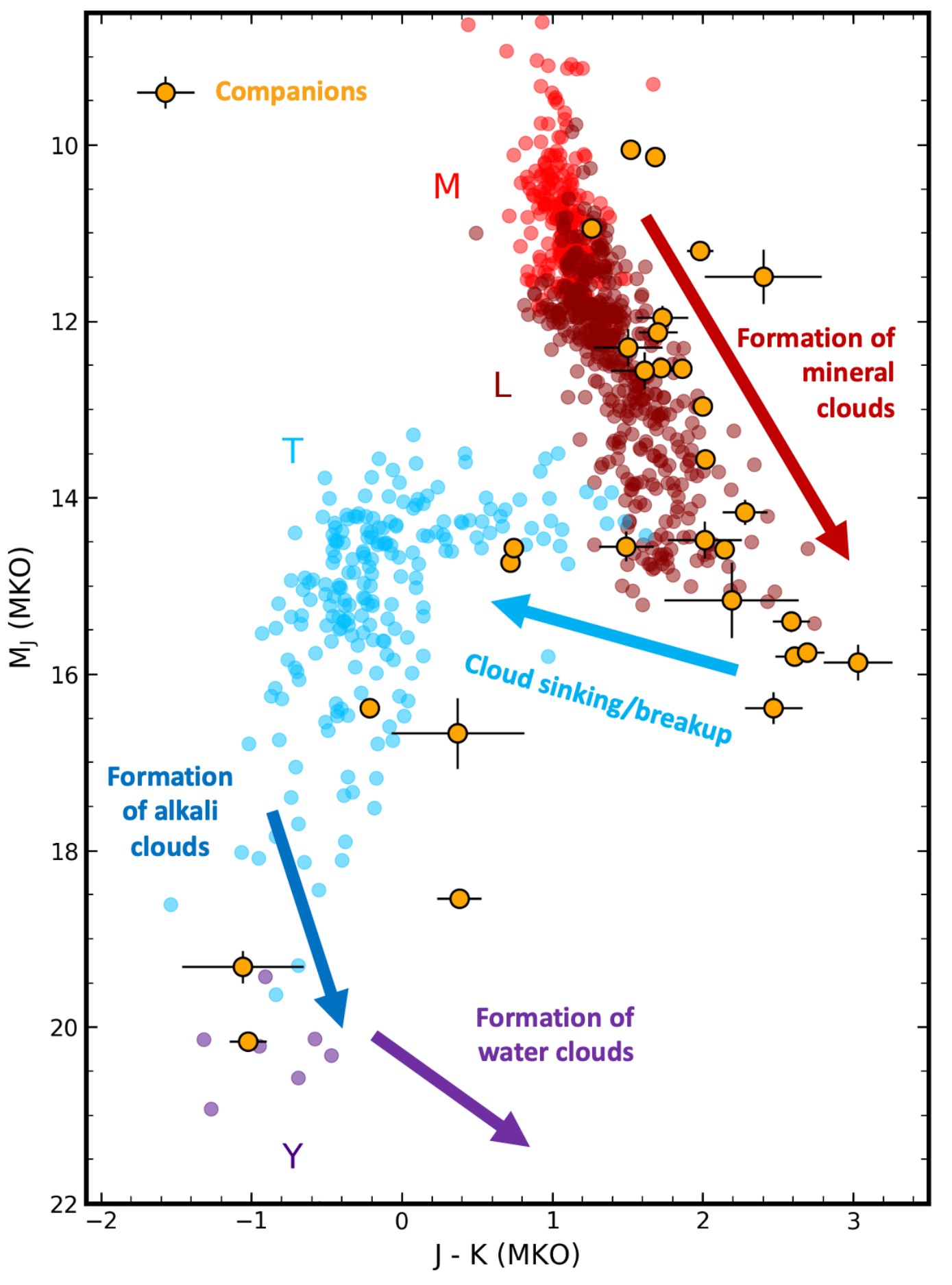}
\caption{Color-magnitude diagram of field dwarfs (M dwarfs: red; L dwarfs: dark red; T dwarf: blue; Y dwarfs: indigo) and directly imaged companions (orange). Here we only include companions that may be exoplanets, which we take to be objects indicated by ``Y'', ``Y?'', and ``N?'' in the ``exoplanet'' column of \citeA{best2020zenodo}, along with objects with those designations that are part of binaries. We also include VHS J125601.92-125723.9 b and SDSSJ224953.46+004404.6A, which were stated to possess masses near the deuterium burning limit by \citeA{bowler2016}. Data are taken directly from \citeA{best2020zenodo}, which has been compiled by \citeA{dupuy2012,dupuy2013,liu2016,best2018,best2020}. Only objects with near-infrared photometry available in MKO magnitudes are included. Annotations indicate our current understanding of cloud evolution on brown dwarfs. }
\label{fig:emission}
\end{figure}

%compared to the toy model trends in Equation \ref{eq:toymod} (dashed lines)

\subsubsection{The Brown Dwarfs Legacy}

The large semi-major axes of directly imaged planets allow for analogies to be drawn between them and isolated field brown dwarfs, for which we have observations of higher quality and quantity \cite<e.g.,>{helling2014,liu2016,line2017}. As Figure \ref{fig:emission} shows, directly imaged companions and field brown dwarfs are similar in their near-infrared colors and luminosities, which in turn are controlled by the formation and evolution of clouds as brown dwarfs age \cite{kirkpatrick2005,lodders2006,bailey2014}: the M-L transition is characterized by the formation of titanium and vanadium clouds, removing TiO and VO gas absorption from the spectrum \cite{lodders1999,burrows1999,lodders2002tiv}. The reddening of the L dwarfs with decreasing luminosity is likely due to increasing cloud optical depth with the formation of silicate and iron clouds \cite{allard2001,marley2002,tsuji2002}. {The significant spread in J-K colors of L dwarfs have been attributed to variations in metallicity and gravity, though the existence of high altitude aerosol layers composed of submicron particles in addition to the mineral clouds have been suggested for the reddest field L dwarfs \cite{hiranaka2016}.} The departure of objects towards bluer near-infrared colors at the L-T transition is partly due to increased methane absorption, but also the sinking of the clouds below the photosphere and/or breaking up of the clouds \cite{ackerman2001,burgasser2002,knapp2004,tsuji2003,stephens2009,marley2010}, though non-cloud explanations have also been proposed \cite{tremblin2016}. The dimming and reddening of late T dwarfs is thought to be due to condensation of sulfides and chlorides \cite{morley2012,line2017,zalesky2019}, while the transition to Y dwarfs occurs with the appearance of ammonia gas absorption, {followed by water condensation for the coolest Y dwarfs discovered so far  \cite{lodders2002a,burrows2003,hubeny2007,cushing2011,morley2014,leggett2015,morley2018}}. Similar condensation sequences and chemical transitions should occur in the atmospheres of directly imaged planets as they evolve. 

In addition to luminosity and color variations over cosmic timescales, brown dwarfs also exhibit temporal variability in broadband emission and spectra as they rotate. This observed variability is indicative of heterogeneous aerosol distributions, including holes in aerosol layers, multiple layers, and variable layer thicknesses, with a possible higher concentration of variable objects at the L-T transition \cite{artigau2009,radigan2012,biller2013,heinze2013,radigan2014,radigan2014rare,crossfield2014,faherty2014,burgasser2014,wilson2014bd,buenzli2014,buenzli2015feh,cushing2016,biller2017,artigau2018,eriksson2019,lew2020,vos2020}. This points to cloud breakup as potentially contributing to the increasingly blue near-infrared colors of later spectral types. Furthermore, differences in variability amplitudes between atmospheric windows and absorption features can reveal the location of the aerosol layers. For example, the discovery that variability amplitudes in spectral windows are larger than those in wavelength-adjacent absorption features in several L-T transition objects shows that an aerosol layer with spatially variable thickness likely exists between the optical depth unity altitudes (where optical depth equals 1) at these two wavelengths \cite{apai2013,buenzli2015,buenzli2015feh}. Conversely, the variability amplitudes of some mid-L dwarfs are linear with wavelength across absorption bands, which is suggestive of an aerosol layer at high altitudes above the optical depth unity altitudes of the absorbers \cite{yang2015,lew2016,schlawin2017}. This in turn implies that the sinking of aerosol layers also occurs as L dwarfs transition to T dwarfs. Complicating these analyses is the recognition that several brown dwarfs' light curves are aperiodic, indicating weather-like processes where aerosol distributions change within a timescale comparable to the rotation period, typically ranging from several hours to several days  {\cite{artigau2009,radigan2012,gillon2013,yang2016,karalidi2016,apai2017,tan2019,apai2021}}. 

\subsubsection{Directly Imaged Planets}

Near-infrared spectroscopy of directly imaged companions has shown that, like isolated brown dwarfs, aerosols are common in their atmospheres and that the distribution of aerosols appears to be heterogeneous, with some objects exhibiting temporal variability in photometry and spectra \cite{marois2008,currie2011,skemer2012,marley2012,bonnefoy2013,skemer2014,ingraham2014,macintosh2015,zhou2016,bonnefoy2016,rajan2017,samland2017,delorme2017,greenbaum2018,muller2018,biller2018rev,manjavacas2019,lew2020,bowler2020,zhou2020,wang2020}. Trends in wavelength-dependent variability with spectral type are also seen among companions, where L-type objects {(those with spectra similar to L dwarfs)} tend to have gray or linear dependence while L-T transition objects show lower variability amplitudes in absorption features \cite{manjavacas2018,zhou2018,milespaez2019}. Importantly, these observations, along with those of young, low gravity, isolated objects \cite<e.g.>{metchev2015,biller2015,gizis2015,liu2016,faherty2016,biller2018,vos2019,vos2020}, show that lower gravity objects tend to be more variable and possess redder near-infrared colors (higher J-K) compared to higher gravity objects of the same effective temperature (Figure \ref{fig:emission}). 

\subsubsection{Transiting Exoplanets}

While many transiting hot and warm Jupiters exhibit similar atmospheric temperatures as isolated brown dwarfs and wide orbit companions, the intense stellar irradiation that they experience while tidally locked to their host stars mean that they possess fundamentally different atmospheric thermal structures both vertically and horizontally. These differences have been revealed by observations of thermal emission from their permanent daysides \cite<e.g.>{deming2005,charbonneau2005,deming2007,charbonneau2008,knutson2008,stevenson2010,majeau2012,dewit2012,kreidberg2014w43b,arcangeli2018,evans2019,wallack2019,garhart2020} and phase curves \cite<e.g.>{knutson2007,knutson2012,cowan2012,lewis2013,zellem2014,stevenson2014pc,wong2015exo,demory2016,wong2016exo,kreidberg2018pcurve,kreidberg2018pc43b,kreidberg2019}. Meanwhile, complementary observations of reflected light in the optical probe the longitudinal distribution of aerosols and the dayside albedo, as controlled by the reflectivity of aerosols and molecular absorption \cite<e.g.>{rowe2008,borucki2009,snellen2009,kipping2011,esteves2013,shporer2014,shporer2019,wong2020tess,bourrier2020,beatty2020,vonessen2020,jansen2020}. We refer the reader to \citeA{parmentier2018,alonso2018} for comprehensive reviews.

\begin{figure}[hbt!]
\centering
\includegraphics[width=0.7 \textwidth]{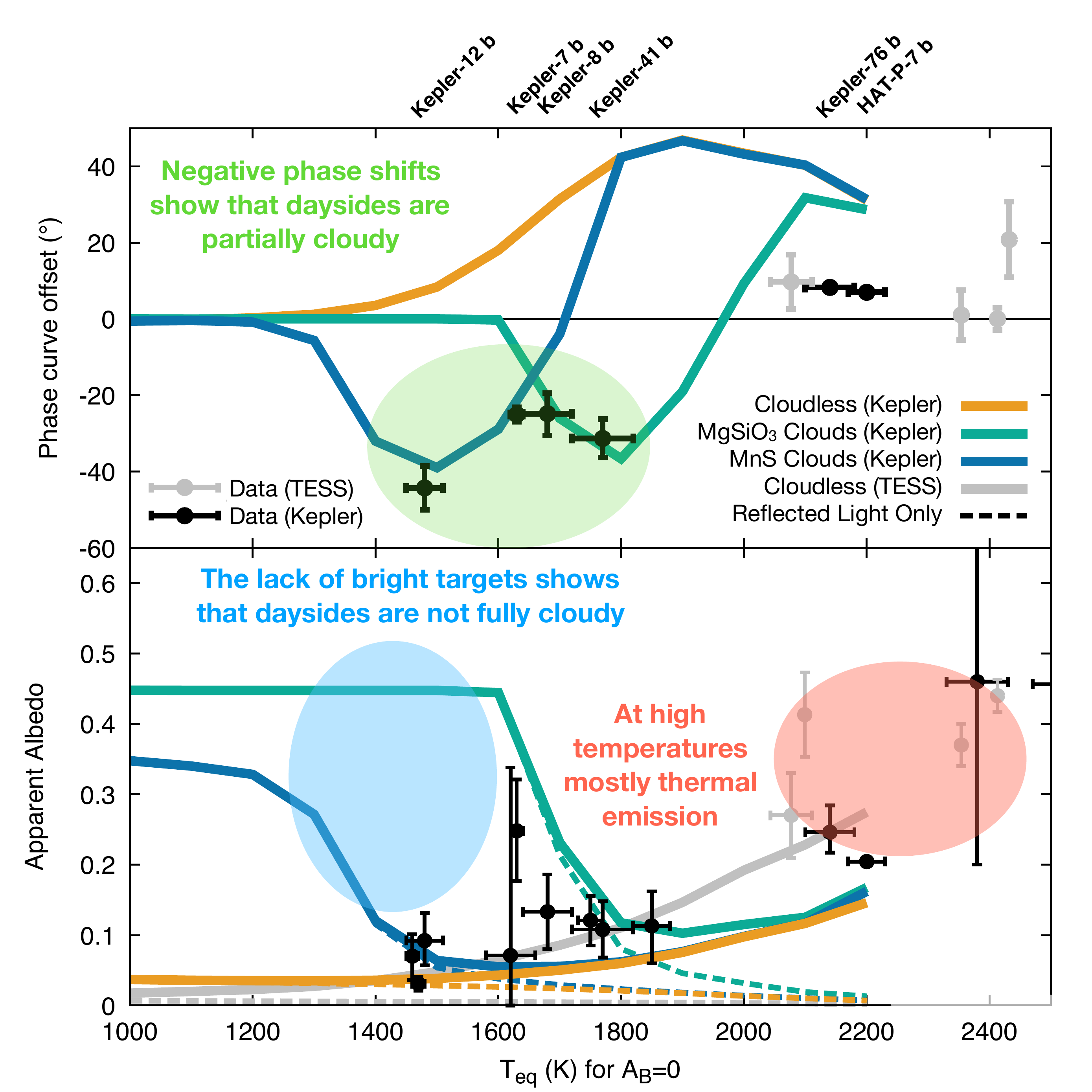}
\caption{Optical phase curve offsets (top) and apparent albedos (bottom) of giant exoplanets in the Kepler (black points) and TESS (gray points) bandpasses compared to global circulation models assuming cloudless atmospheres (orange and gray curves for Kepler and TESS bandpasses, respectively) and atmospheres post-processed with MgSiO$_3$ (green curve) or MnS (blue curve) clouds. The apparent albedo includes both reflected and emitted light; the reflected light-only albedo is shown in the dashed curves, indicating that most of the photons received from the daysides of ultra-hot giant exoplanets are emitted rather than reflected. Two planets are not shown here, as they are situated beyond the plot limits: KELT-1b, which has an apparent albedo of 0.7 in the TESS bandpass \cite{vonessen2020}, and WASP-100b, which has a controversial hot spot measurement in the TESS bandpass (see \citeA{jansen2020} vs. \citeA{wong2020tess}). The figure is updated from \citeA{parmentier2016}. {\copyright~AAS. Reproduced with permission.}}
\label{fig:pcurves}
\end{figure}

Combined optical and infrared observations have revealed significant longitudinal heterogeneity in the distribution of aerosols in transiting exoplanet atmospheres. For example, dayside near-infrared photometry and spectra can be explained without the need for optically thick aerosols down to the pressure levels probed, suggesting either a lack of aerosols altogether or that aerosols form at pressures higher than the planets' photosphere  \cite{lee2012,line2014,barstow2014,kataria2015}. Similarly, the observed optical geometric albedoes are low ($\leq$0.1) for nearly all giant transiting planets \cite<Figure \ref{fig:pcurves};>{coughlin2012,heng2013,angerhausen2015,bell2017,dai2017,mocnik2018,niraula2018,mallonn2019,kane2020}, consistent with significant gas absorption without substantial reflective aerosols \cite{marley1999exo,sudarsky2000,seager2000curve}. One outlier, Kepler-7b, which possesses a geometric albedo $\sim$0.3 \cite{demory2011}, also exhibits a maximum in its optical phase curve west of the substellar point; this has been interpreted as the presence of highly reflective aerosols covering a small fraction of the western limb \cite{demory2013,garciamunoz2015,webber2015}. Westward-shifted maxima in optical phase curves have been observed for several objects (Figure \ref{fig:pcurves}), which all have $T_{eq}$ $<$2000 K \cite{esteves2015,shporer2015}, suggesting that reflective aerosols may be common on the western limbs of hot and warm giant exoplanets, even though most of the dayside hemisphere may be clear. HD 189733b also exhibits a high albedo of 0.4$\pm$0.12 at blue optical wavelengths, decreasing to $<$0.12 in the red optical \cite{evans2013}, but this can be explained by Rayleigh scattering by atmospheric gas molecules alone \cite{barstow2014}. 

Several recent estimations of geometric albedoes in the \textit{TESS} bandpass have yielded high values ($>$0.2) for ultra-hot ($T_{eq}$ $>$ 2000 K) giant exoplanets \cite<Figure \ref{fig:pcurves};>{wong2020tess,vonessen2020}. However, because the measured flux in the TESS bandpass for ultra hot Jupiters is dominated by thermal emission rather than reflected light, the estimated albedoes are very sensitive to the assumptions made when estimating the thermal contamination. Assumptions about the stellar ellispsoidal effects \cite{shporer2017,wong2020tess}, chemical profiles \cite{parmentier2018ultra}, thermal profiles \cite{lothringer2018heat}, heat redistribution efficiency \cite{arcangeli2019}, or the lack of 2D effects \cite{taylor2020} can all lead to an underestimation of the thermal emission contamination and thus an overestimation of the actual albedo. Cooler planets observed at shorter wavelenghths \cite{shporer2015} are much more likely to yield precise albedo measurements. Finally, the geometric albedos of hot and warm ($T_{eq}$ $\geq$ 600 K) Neptune-size, mini-Neptune, and rocky exoplanets in the \textit{Kepler} bandpass have been constrained to low values ($\leq$0.2) \cite{demory2014,sheets2014,sheets2017,jansen2018}. {Though the large measurement uncertainties have prevented solid conclusions to be drawn about aerosol properties in their atmospheres, reflective, spatially extensive aerosols are unlikely.}

\begin{figure}[hbt!]
\centering
\includegraphics[width=0.8 \textwidth]{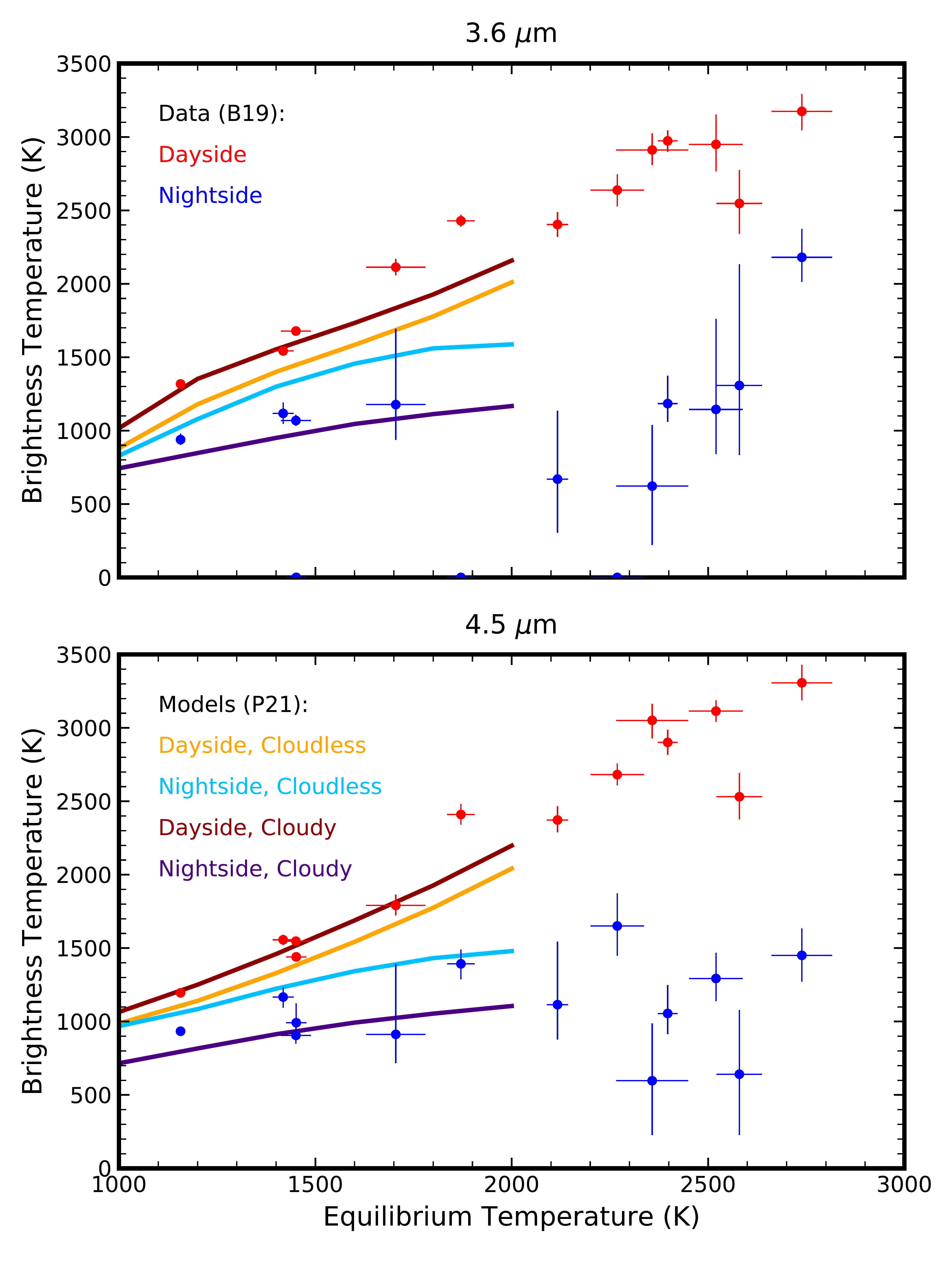}
\caption{{Observed brightness temperatures of the daysides (red) and nightsides (blue) of hot Jupiters at the 3.6 (top) and 4.5 $\mu$m (bottom) \textit{Spitzer} bands from \citeA{beatty2019}, which probe atmospheric temperature and opacity structures, including the effect of aerosols. Brightness temperatures computed by \citeA{parmentier2021} using a GCM for cloudless hot Jupiters (daysides: orange, nightsides: light blue) and hot Jupiters with cloudy nightsides (daysides: dark red, nightside: indigo) are shown for comparison.}}
\label{fig:emissiondaynight}
\end{figure}

The nightsides of transiting giant exoplanets, as probed by thermal phase curves, show an intriguing pattern. Multiple studies \cite{beatty2019,keating2019} have now shown that the nightside brightness temperature, as measured by \textit{Spitzer}, is a nearly constant $\sim$1100 K, while the dayside brightness temperature increases linearly with $T_{eq}$ (Figure \ref{fig:emissiondaynight}). This pattern persists up to at least $T_{eq}$ $\sim$ 2500 K. One possible explanation for this phenomenon is that observations of the nightside at near-infrared wavelengths probe the top of a cloud layer that persists on all transiting giant exoplanets with $T_{eq}$ $\leq$ 2500 K, such that the nightside brightness temperature is tied to the condensation temperature of that cloud \cite{beatty2019}. {Indeed, current efforts to interpret the amplitudes and phase shifts of hot Jupiter thermal phase curves using non-aerosol explanations, such as higher metallicity \cite{kataria2015,drummond2018met} and disequilibrium chemistry \cite{drummond2018dis,mendonca2018,steinrueck2019}, have met limited success.} 

%above which the nightside brightness temperature begins increasing with increasing $T_{eq}$

%\begin{equation}
%\label{eq:toymod}
%T_{day}^4 = 2T_{eq}^4 - T_{night}^4
%\end{equation}

%\noindent where $T_{night}$ = 1100 K

In summary, probes of exoplanet atmospheres by multiple observational methods have revealed an abundance of aerosols across all varieties of worlds. By taking advantage of the growing number of exoplanets amenable to characterization, we have found hints of trends in aerosol distribution with planet temperature and discovered the spatial heterogeneity of aerosols. Furthermore, we have shown through our discussions of both directly imaged exoplanets and transiting exoplanets that the level of stellar irradiation and orbital distance greatly affects aerosol distributions. In the next section, we overview the theoretical tools that have been brought to bear to explain our diverse observations.

%smoran14: I think a mini-paragraph here that briefly zooms back out and gives like 2 big picture takeaways from observations could be really helpful for the "give solid conclusions" goal of the review. Same thing for theory. Lab section might be short enough that it's not needed.

%Hannah Wakeford: I agree there needs to be a bit more of a punch here - what should I take away from the comparison of brown dwarfs and transiting planets?

%by several studies. \citeA{sheets2014,sheets2017} averaged \textit{Kepler} light curves on a population scale and found values of $\sim$0.2 for sub-Saturns/exo-Neptunes and $\leq$0.1 for smaller planets, though uncertainties are relatively large ($>$25\%). \citeA{demory2014} computed a population-level geometric albedo of 0.16-0.3 for planets with radii $<$ 2 $\times$ the Earth radius, higher than those of \citeA{sheets2017}, possibly due to sampling bias. Finally, \citeA{jansen2018} provided upper limits to the albedoes of Neptune-size and rocky planets in the \textit{Kepler} bandpass of $<$0.23 and $<$0.42, respectively, at 95\% confidence. The low albedos and high temperatures of the rocky planets rule out a bright, Venus-like sulfuric acid cloud layer and may be due to seeing an absorptive surface directly \cite<e.g.>{hu2012,kreidberg2019}. The Neptune-size and mini-Neptunes' albedoes are within 1$\sigma$ of the albedoes of giant exoplanets, providing little evidence of drastically different dayside aerosol properties. 

\section{Insights from Theory}\label{sec:theo}

The formation, evolution, and spatial and size distribution of aerosols depend on interactions between the atmospheric thermal structure, wind patterns, and microphysical processes \cite{pruppacher1978}. For clouds, these processes include nucleation, the conversion of condensate vapor into solid or liquid either directly (homogeneous) or with the aid of a foreign surface (heterogeneous) often in the form of a ``condensation nuclei''; condensation, the growth of cloud particles through the uptake of vapor; coagulation, the growth of cloud particles through collision and sticking; evaporation, the decrease in particle mass due to loss of condensate molecules to the atmosphere; and transport by sedimentation, diffusion, and advection by winds. For hazes under our definition (${\S}$\ref{sec:def}), formation and growth through chemical reactions and coagulation, transport, {and loss through thermal decomposition, nucleation (i.e. acting as condensation nuclei for clouds), and wet deposition} are the primary processes. In this section, we review the extent to which current exoplanet models capture the aforementioned physical processes and what they can tell us about what we have observed. 

%The spatial and size distribution of aerosols in exoplanet atmospheres are controlled by a combination of these microphysical processes and atmospheric properties like temperature, composition, and dynamics.

\subsection{Modelling Exoplanet Aerosols}\label{sec:mod}

\subsubsection{Microphysical and Parametrized Models}\label{sec:modeldiscussion}

Exoplanet aerosol models span a continuum in complexity, from single-variable parameterizations to computationally expensive simulations of aerosol microphysics in 3D, with each type of model serving a different purpose. Highly parameterized models are typically used for retrieval studies where rapid model execution is key. These models use a handful of variables (e.g., cloud top pressure) to capture only the first order impacts of aerosols on observations, such as changing/enhancing the spectral slope at optical wavelengths and reducing the amplitude of molecular features in the infrared, without treatment of specific physical processes \cite<e.g.,>{benneke2012,greene2016,line2016hd209,barstow2017,burningham2017,tsiaras2018,goyal2018,mai2019,zhang2019,pinhas2019,molliere2019,barstow2020,barstow2020cloud}. Some retrieval frameworks have included more complex aerosol models that consider {various combinations of:} Mie scattering by spherical particles {\cite{lee2014,zhang2019,benneke2019,lacy2020jwst2}}, aerosol layers that vary with altitude {\cite{lupu2016,zhang2019,benneke2019,damiano2020}}, aerosol composition \cite{lee2014,fisher2018}, and spatial heterogeneity {\cite{line2016,macdonald2017,feng2018}}. These more complex models allow for more physical interpretations of how aerosols affect observations, but they come at a price of a greater number of parameters, some of which may not be well constrained by the data we currently possess \cite<e.g.,>{fisher2018}. In addition, \citeA{mai2019,barstow2020cloud} showed that the retrieved atmospheric temperature and composition from transmission spectra are largely insensitive to the chosen aerosol parameterization, as long as aerosols are not ignored in the retrieval. In contrast, the aerosol properties retrieved using the different parameterizations may be substantially different from each other, suggesting that consistent constraints on exoplanet aerosols may be difficult to obtain through retrievals that use simple aerosol parameterizations. 
%In the last decade, simple, 1D models have been used extensively in retrieval studies and the generation of model atmosphere grids, where the aerosol opacity is defined using a wavelength-independent cloud top pressure, a multiplicative factor on the Rayleigh scattering slope in transmission, and/or a power law   \cite<e.g.>{benneke2012,greene2016,barstow2017,tsiaras2018,goyal2018,zhang2019,pinhas2019,molliere2019,barstow2020}. More detailed aerosol parameterizations have included Mie scattering and/or a fall-off in aerosol opacity with altitude \cite{tsuji2002,burrows2006,lee2014,zhang2019,benneke2019,damiano2020}, a consideration of aerosol composition \cite{tsuji2002,burrows2006,lee2014,fisher2018}, and spatial heterogeneity \cite{macdonald2017}. \citeA{mai2019} showed that the retrieved atmospheric temperature and composition are largely insensitive to the chosen aerosol parameterization, as long as one is included, i.e. ignoring aerosols in retrievals could lead to incorrect results. In contrast, the cloud properties retrieved using the different parameterizations were different from each other, suggesting that constraints on exoplanet aerosols may be difficult to obtain through retrievals alone. 

%One caveat of this study is that the synthetic data used for the retrievals were generated using another cloud model \cite{ackerman2001}, which may not fully capture the full range of exoplanet aerosol distributions.

%CHIMERA, PLATON, Bjorn's model, ATMO, POSEIDON, NEMESIS - see Barstow 2020 for a good review on clouds in retrievals.

\begin{figure}[hbt!]
\centering
\includegraphics[width=0.8\textwidth]{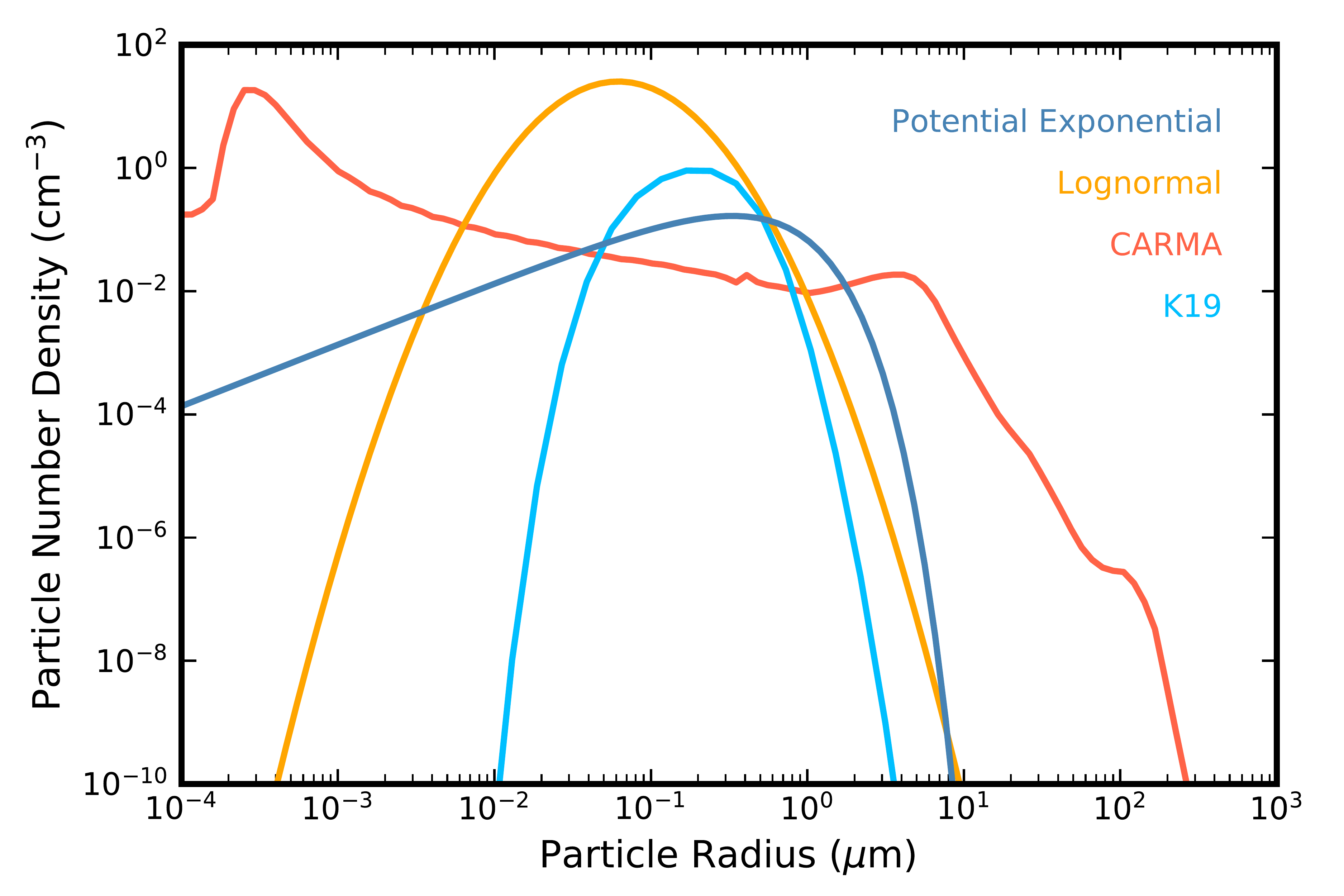}
\caption{Example particle size distributions from aerosol models. Shown are the parameterized potential exponential \cite<dark blue; e.g.>{helling2008b} and lognormal \cite<orange; e.g.>{ackerman2001} functions, along with binned size distributions from the cloud simulations of \citeA{gao2020} using \texttt{CARMA} (red) and the haze simulations of \citeA{kawashima2019b} (K19; light blue).  }
\label{fig:sizedist}
\end{figure}

Aerosol models that include some of the physical processes that control aerosol distributions, but which are still parameterized to be computationally inexpensive are often found as a part of radiative-convective equilibrium models. While these models are more computationally expensive than retrievals, they are useful for generating model grids that elucidate the roles of specific parameters across populations of objects. In contrast, retrievals typically seek to extract physical parameters from observations by rapidly exploring the parameter space for a single object. These more complex aerosol models typically treat aerosol compositions computed from thermochemical equilibrium (see ${\S}$\ref{sec:def}) and assume either a mean particle size or a functional form for the aerosol size distribution (Figure \ref{fig:sizedist}), allowing them to compute aerosol optical properties assuming Mie scattering. The differences between these models are due to how they parameterize aerosol microphysics. The cloud model of \citeA{ackerman2001}, for example, computes cloud distributions by balancing particle sedimentation with vertical mixing, while the vertical extent of the clouds is controlled by a sedimentation efficiency parameter. In contrast, the model of \citeA{cooper2003} computes the mean particle size by balancing the timescales of microphysical processes following \citeA{rossow1978}; free parameters include the supersaturation, which controls the nucleation and condensation timescales, and the sticking coefficient that controls the coagulation timescale. These two approaches both allow for relatively fast computations of profiles of particle sizes, cloud mass mixing ratios, and cloud optical properties. While these models' reliance on tunable parameters hinders their predictive powers, it also allows them to explore a large range of cloud properties and how they affect observations. \citeA{hu2012model} describes an alternative aerosol model coupled to a photochemical model where the particle size is a free parameter, sedimentation is treated explicitly, and condensation and evaporation is computed through associated timescales. This model considers cloud compositions produced by photochemical reactions like sulfuric acid and sulfur (${\S}$\ref{sec:def}) and has been used mostly for terrestrial planet atmospheres thus far.

 \begin{table}[hbt!]
 \caption{Properties of microphysical models of exoplanet aerosols}
 \centering
 \begin{tabular}{l c c c c c c c}
 \hline
  Model & Moment/Bin & Mixed/Pure & C/H  & Nucleation & Haze Formation & Transport & Reference \\
 \hline
   \texttt{DRIFT} & Moment & Mixed & C & Comp. & \nodata &  Relaxation & [1]  \\
   O17  & Moment & Pure & C\&H & Hybrid & Param. & Advection & [2]  \\
   L17 & Bin & Pure & H  & \nodata & Hybrid & Diffusion &[3] \\
   \texttt{CARMA} & Bin & Mixed & C\&H & Comp. & Param. &Diffusion &[4] \\
   K18 & Bin & Pure & H  & \nodata & Photo. &Diffusion & [5]\\
   \texttt{ARCiS} & Moment & Pure & C & Param. & \nodata  &Diffusion & [6] \\
 \hline
 %\multicolumn{6}{r}{$^{a}$[1] \citeA{helling2008drift}; [2] \citeA{ohno2017}; [3] \citeA{lavvas2017}; [4] \citeA{gao2018a}; [5] \citeA{kawashima2018}; [6] \citeA{ormel2019} }
 %\multicolumn{6}{l}{$^{b}$The models described here include }
 \end{tabular}
 [1] \citeA{helling2008drift}; [2] \citeA{ohno2017}; [3] \citeA{lavvas2017}; [4] \citeA{gao2018a}; [5] \citeA{kawashima2018}; [6] \citeA{ormel2019}
 \label{tab:micromodels}
 \end{table}

The most complex 1D aerosol models treat microphysical processes kinetically, producing particle distributions by balancing the rates of the individual processes. These models are used for exploring how different aerosol processes interact with each other and how aerosol distributions form and evolve. Table \ref{tab:micromodels} shows how current kinetic models compare on several modeling techniques. Some of the models parameterize the particle size distribution using moments of that distribution, such that the actual shape of the distribution must be user-prescribed, while other models are able to resolve the size distribution using mass bins. Parameterizing the size distribution saves computation power, but may fail to capture multiple particle size modes (Figure \ref{fig:sizedist}), leading to significant differences in aerosol opacity {and its wavelength dependence} \cite{powell2019}. The models also differ in whether they treat particles as being composed of a single composition or a mixture of multiple compositions. Given the large number of potential condensates at high temperatures (Figure \ref{fig:cloudcomp}), the existence of mixed composition particles is likely. \texttt{DRIFT} {(not an acronym)} and \texttt{CARMA} {(Community Aerosol and Radiation Model for Atmospheres)} treat mixed particles differently, however: \texttt{DRIFT} allows multiple species to condense onto the same cloud particle at the same time via numerous thermochemical reactions, forming well-mixed ``dirty grains'' \cite{helling2006,helling2008b}, a procedure originating from models of mixed dust grains in stellar winds \cite{gail1984,gail1988}. In contrast, \texttt{CARMA} considers only layered particles, such that only the outer most layer can grow by condensation at any one time \cite{gao2018b}. This strategy originates from \texttt{CARMA}'s roots in Earth science \cite{turco1979,toon1979} where condensing water vapor can completely envelope a condensation nucleus, but may not suffice for high temperature exoplanet clouds where multiple condensates may interact.  

%This leads to differences in cloud optical properties and cloud vertical structure. 

The considered kinetic models can be further divided between cloud models (C), haze models (H), and models capable of simulating both types of aerosols (C\&H). Of the models capable of simulating clouds, an important attribute to consider is how they treat nucleation, as that ultimately controls the depletion of condensate vapor and the cloud particle number density and size. The \texttt{ARCiS} {(ARtful modeling Code for exoplanet Science)} framework \cite{ormel2019} parameterizes (Param. in Table \ref{tab:micromodels}) their particle nucleation rate with a Gaussian profile and a user-defined column rate. The model of \citeA{ohno2017} uses a hybrid approach where the heterogeneous nucleation rate is computed based on user-input number densities and sizes of condensation nuclei. \texttt{DRIFT} \cite{helling2008drift} and \texttt{CARMA} \cite{gao2018a} both compute (Comp. in Table \ref{tab:micromodels}) homogeneous nucleation rates of condensation nuclei from classical nucleation theory (or modified versions of it); for subsequent condensation of other cloud compositions the former model considers grain chemistry while the latter model computes heterogeneous nucleation rates. While consideration of nucleation theory is more physical, it has been shown to differ from experimentally determined rates by orders of magnitude for some substances \cite<see e.g.>{oxtoby1992,anisimov2009} and relies on material properties that may not have been measured at the appropriate temperatures, e.g. surface energies \cite{gao2020}. 

Likewise, of the models capable of simulating hazes, a major source of uncertainty is how they compute the haze formation rate profile in the atmosphere, which ultimately determines the haze opacity. While no model has been able to fully simulate the chemical network from simple parent molecule to aerosol particles, some are more parameterized than others. \texttt{CARMA} and the \citeA{ohno2017} model \cite<as described in>{ohno2020} both fully parameterize (Param. in Table \ref{tab:micromodels}) haze formation rates through user-chosen production rate profiles and initial particle sizes. The model of \citeA{lavvas2017} is similar except the column rate is computed from photochemical models (Hybrid in Table \ref{tab:micromodels}) under the assumption that some percentage of the pertinent photochemical reactions (a ``haze formation efficiency''), typically involving the photolysis of hydrocarbons and nitriles, lead to haze formation. {It is also the only model thus far to explicitly treat thermal decomposition of haze particles.} The model of \citeA{kawashima2018} takes this a step further by equating the production rate profile to some percentage of the rate profiles of the chosen photochemical reactions (Photo. in Table \ref{tab:micromodels}), though the initial particle size is still a free parameter. Similar strategies have also been adopted by other studies \cite{morley2013,morley2015,zahnle2016} that don't consider kinetic models of aerosol microphysics. In addition, several models have simulated exoplanet haze particles as fractal aggregates \cite{arney2017,arney2018,adams2019,lavvas2019}, much like haze particles on Titan {\cite{lavvas2011agg}}. Several cloud models have also considered aggregates composed of condensates \cite{ohno2020agg,samra2020}.

\subsubsection{Thermal Structure}

The formation of clouds is intimately linked to the thermal structure of the atmosphere. For a given cloud species, too high of a temperature can prevent nucleation and condensation while too low of a temperature might shift cloud formation to deeper layers of the atmosphere that cannot be probed by current observations. 

The thermal structure of exoplanets can be either calculated \textit{a priori} using radiative-convective equilibrium models or retrieved directly from the planets' emission spectra. Radiative-convective models are often used in objects that experience no or little irradiation such as brown dwarfs and directly imaged planets where horizontal advection of heat is not significant. The radiative convective equilibrium profile can be further refined by using a combination of cloud and cloud-free atmospheric patches \cite<e.g.,>{marley2010,morley2014patchy}. Parameterized thermal structures are often used for irradiated planets such as hot Jupiters \cite{guillot2010,heng2012,parmentier2014}, where the presence of trace species that are difficult to characterize, such as metal oxides or metal hydrides, can lead to large uncertainty in the expected thermal structures \cite{fortney2008,gandhi2019}. Furthermore, local radiative-convective equilibrium does not hold for specific parts of the planet, such as the dayside and the limb that are probed by transmission and emission spectra, {as their local thermal structure is determined by the global atmospheric circulation.}

Aerosols have two main effects on the thermal structure of an atmosphere. First, they change the heat transport within the atmosphere: they warm up the atmosphere below the cloud by their increased opacity and cool down the atmosphere above the cloud top by efficiently radiating away heat. Second, aerosols change the albedo and thus the emissivity of the planet. The change in albedo leads to a change in the total energy received by the atmosphere and thus a change in the mean thermal structure of the planet. The change in emissivity leads to a change in the ability of the atmosphere to re-emit light and thus a change in the relationship between thermal structure and observed spectra \cite{lee2013,lavie2017,burningham2017,molliere2020}. 

\subsubsection{Atmospheric Transport}

Atmospheric transport is needed both for aerosols to stay aloft in the atmosphere and for fresh gaseous species to replenish the depleted gas in the aerosol formation region. If no vertical mixing were present, all condensable species would rain out of the visible atmosphere. In 1D models, vertical mixing is usually assumed to be diffusive in nature and parameterized by a {eddy diffusion coefficient} $K_{zz}$, which is highly uncertain. {As a diffusion coefficient, $K_{zz}$ has units of length-squared per unit time. The ``zz'' subscript denotes motion in the z, i.e. vertical direction. In addition to being used in aerosol models to transport particles and condensate vapor, $K_{zz}$ has also been frequently used in chemical kinetics/photochemistry models to approximate transport of gases \cite<e.g.>{moses2016}.} {$K_{zz}$ is an approximation of all large scale transport in a planetary atmosphere, including atmospheric circulation, gravity waves, and convection, most of which cannot be explicitly represented in 1D models. As some of these processes are not actually diffusive \cite{zhang2018a,zhang2018b}, the use of $K_{zz}$ to represent atmospheric transport and how its profile in the atmosphere is calculated require caution.} For objects that are mainly convective, mixing length theory is often used to estimate the $K_{zz}$ profile \cite<e.g.,>{gierasch1985,ackerman2001}. This is not valid for atmospheres that are predominantly radiative, such as those of transiting exoplanets. 

On tidally locked planets, advection by the atmospheric circulation is likely the main source of vertical mixing, as opposed to turbulence or wave breaking. As first shown by \citeA{parmentier2013} using passive tracers in a 3D global circulation model (GCM), the mean vertical transport of particles in the atmospheres of hot Jupiters is surprisingly well represented by a 1D diffusion approach, {resulting in a $K_{zz}$ profile that increases with decreasing atmospheric pressure as a power law}. The magnitude of $K_{zz}$ is, however, a hundred times smaller than would be expected from extrapolating the mixing length parameterization or by multiplying the root mean square of the vertical velocities by the atmospheric scale height. \citeA{zhang2018a,zhang2018b} explored a wider range of atmospheric circulation patterns and showed that the $K_{zz}$ used in 1D models should be different for different chemical and aerosol species. They further identified specific cases, such as when photochemical hazes form in the upper layers of a dayside updraft, which would require a negative $K_{zz}$, representing local concentration rather than dilution. Lastly, \citeA{komacek2019} proposed an analytical formula for $K_{zz}$ that is based on the Earth stratosphere framework developed by \citeA{holton1986}; they showed that $K_{zz}$ should depend on both the strength of the circulation and the timescale at which a given species is lost. Their formalism, however, was developed for gaseous chemical species only, which are not conserved. As such, it is not yet clear how to adapt it to aerosols that are usually conserved when settling vertically. In the deeper atmosphere, other processes likely start to dominate vertical mixing such as wave breaking \cite{fromang2016} or shear instability driven turbulence \cite{menou2019}.  

Not all 1D aerosol models treat transport as a diffusive process. The 1D version of \texttt{DRIFT} considers a newtonian relaxation scheme for the chemical abundances instead of solving an actual diffusion equation \cite{woitke2004}, and assumes that the particles are fully decoupled from the flow. In every atmospheric layer the gaseous composition relaxes towards the initial conditions. Though this approach is more straightforward numerically, it can lead to orders of magnitude differences in the cloud distribution compared to a model that uses a diffusion approach \cite{woitke2020}. \citeA{ohno2017} also does not consider diffusion; instead, chemical species and particles are lifted upward by a constant advection along the 1D column. This approach is correct when modelling cloud formation in an updraft, but might be incorrect when modelling the spatially averaged atmosphere, where all updrafts are compensated by downdrafts. 

% these estimates to predominantly radiative atmospheres requires adjustements as discussed in~\citet{Lines}. 

\subsubsection{3D Aerosol Models}

Aerosol models of various complexities have been incorporated into GCMs in an effort to understand the global aerosol distribution on exoplanets. On the more parameterized end, cloud particles have been treated as radiatively passive \cite{parmentier2013,charnay2015a,komacek2019} and active \cite{charnay2015b} tracers that typically have a user-defined particle size distribution, are advected by the circulation, and may be removed through a parameterization of condensation and sedimentation. This technique has been useful in revealing how aerosols are transported in an atmosphere, particularly whether they can be lofted to high altitudes to explain muted gas spectral features. Alternatively, parameterized cloud distributions are prescribed onto the 3D grid of the GCM based on observations \cite{roman2017} or as 1D columns in which cloud formation is evaluated based on whether condensate vapor is locally supersaturated without advection of the clouds {\cite{parmentier2016,tan2017,parmentier2018ultra,roman2019,harada2019,tan2020,parmentier2021,roman2021}}. More complex 1D cloud models, like \texttt{DRIFT} \cite{lee2015,helling2016,helling2019,helling2019hat,helling2020} and the \citeA{ackerman2001} model \cite{lines2019} have also been incorporated into GCMs in this fashion. Both radiatively active and post-processed clouds (i.e. clouds added to the model atmosphere after a cloud-less GCM converged) have been considered. These studies have investigated how aerosols affect a planet's thermal emission and albedo, particle heating and cooling, and the impact of local aerosol formation on gas abundances. Several studies have more fully coupled \texttt{DRIFT} to GCMs \cite{lee2016,lee2017,lines2018,lines2018neph}, such that both particle advection, cloud microphysics, and cloud radiative feedback are considered simultaneously. While these models capture more of the interactions between the different physical processes, running them until all modeled processes can converge is currently computationally prohibitive.  

\subsection{Aerosol Model Predictions and Comparisons to Data}\label{sec:micro}

Exoplanet aerosol models have been used to interpret a variety of observations of exoplanet atmospheres and also predict future observations. In particular, many studies have focused on explaining observations of individual planets with aerosol models, either as part of retrieval frameworks \cite<e.g.>{kreidberg2014,macdonald2017,wakeford2018,benneke2019,molliere2020} and/or more complex forward models \cite<e.g.>{fortney2005,barman2011,marley2012,bonnefoy2013,lee2015,rajan2017,chachan2019}. These studies have revealed a diversity of exoplanet atmospheres across planetary parameter space. However, due to limited data these comparisons often run into degeneracies and it is unknown how applicable their conclusions are to all exoplanets. Therefore, in this section we will mostly focus on studies that have attempted to explain or predict how aerosols impact whole populations of exoplanets, though we will also discuss several benchmark objects. 

\subsubsection{Transiting Exoplanets}

As reviewed in ${\S}$\ref{sec:obs}, the proliferation of exoplanet transmission spectroscopy, emission photometry, and optical and infrared phase curves allow us to probe the vertical and horizontal extent of aerosols in exoplanet atmospheres across a wide range of planetary parameters. These efforts have yielded several important clues on how aerosol distributions vary with planet $T_{eq}$ and longitude: (1) The daysides of giant transiting exoplanets are likely clear while the nightsides and western limbs likely host optically thick aerosols and (2) the vertical extent of aerosols at the limbs, as probed by transmission spectroscopy, may correlate with planet temperature. Several modeling studies have tried to explain these observations. 

\begin{figure}[hbt!]
\centering
\includegraphics[width=0.6 \textwidth]{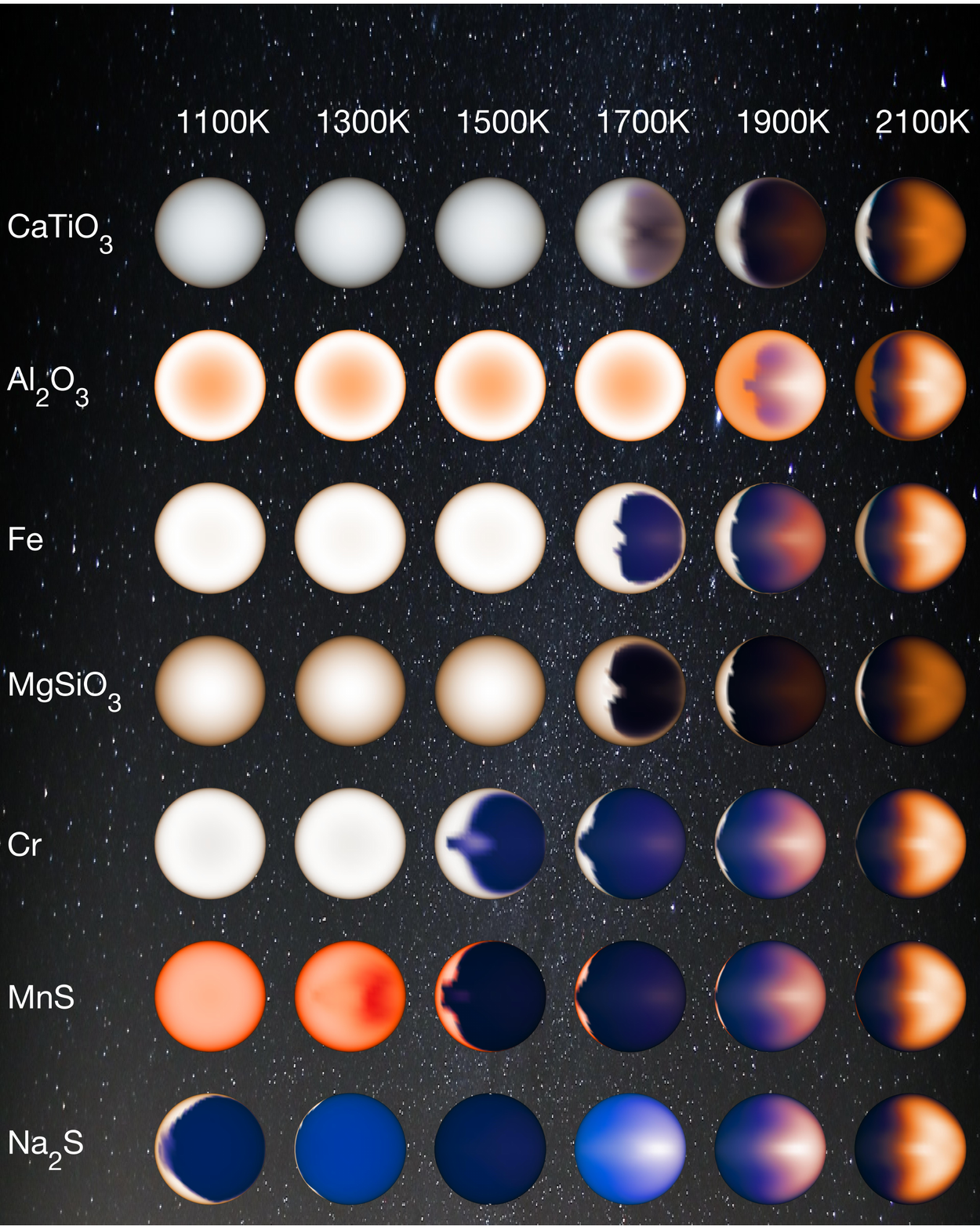}
\caption{Visual appearance of the daysides of a set of exoplanet global circulation models with post-processed clouds \cite{parmentier2016}, {generated with the same approach as those of \citeA{harre2021}}. Each column represents planets of a single equilibrium temperature while each row shows different cloud species. For low equilibrium temperatures, the dayside is often fully cloudy and the color of the cloud depends on its absorption bands in the optical. For intermediate temperatures, the cloud coverage is concentrated on the western part of the dayside, where the planet is cooler. The eastern part, dominated by the hot spot, is cloudless and has a dark blue appearance due to alkali absorption. At high equilibrium temperatures, the daysides are mostly cloudless, with their visual appearance dominated by the thermal emission of the hot spot. {Image credit: NASA/JPL-Caltech/University of Arizona/V. Parmentier.}}
\label{fig:cldmap}
\end{figure}

\citeA{parmentier2016} computed the total thermal emission and reflected light fluxes in the \textit{Kepler} bandpass of a grid of hot giant exoplanets by adding post-processed, parameterized clouds to the output of a GCM and compared their results to observed optical phase shifts and apparent albedos (Figure \ref{fig:pcurves}). They found that a transition in cloud composition, as determined by local thermal stability of condensates predicted by thermochemistry models, could explain the data: at the highest temperatures, the dayside is devoid of aerosols with the flux dominated by thermal emission, which reaches a maximum towards the east limb; as temperatures decrease, silicate clouds form on the nightside and western limb, where the temperatures are the lowest, and begin extending eastward over the dayside, shifting the brightest longitude in the \textit{Kepler} bandpass to the west and causing reflected light to dominate over thermal emission; at $\sim$1600 K, the observed low albedo necessitates the disappearance of silicate clouds from the dayside, possibly due to sequestration in deeper atmospheric layers; at the same time, MnS clouds replace silicates as the dominant aerosol species on the western limb, perpetuating the westward optical phase curve shift (Figure \ref{fig:cldmap}). \citeA{parmentier2016} also found that cloud radiative feedback causes a net increase in the temperature of the planet due to the greenhouse effect of clouds on the nightside, producing a higher emission flux in the clear parts of the dayside and increasing the day-night temperature contrast, consistent with similar studies with prescribed clouds \cite{roman2019}. {\citeA{parmentier2021,roman2021} extended these works to show that the existence of nightside clouds is a likely explanation for the observed low uniform brightness nightside temperatures of hot Jupiters \cite<Figure
\ref{fig:emissiondaynight};>{beatty2019,keating2019} and thermal phase curve shifts, though the decrease in radiative timescale with increasing equilibrium temperature also strongly contributes.} 

The formation of spatially inhomogeneous clouds could also have a large impact on the spatial distribution of the gaseous species involved in cloud formation. For example, \citeA{helling2019} found that the C/O ratio can vary from sub-solar to super-solar ($\sim$0.3 to $\sim$0.7) due to the evaporation and condensation of oxygen-bearing clouds (e.g. silicates) at different locations around the planet.  

Studies that treat aerosols in GCMs as tracers showed that the atmospheric circulation of hot Jupiters tend to reduce the cloud cover at the equator compared to the mid-latitudes \cite{parmentier2013,komacek2019}. Additionally, when aerosol microphysics is considered \cite{lee2016,lines2018}, latitudinal variations in particle size and composition were predicted, with small particles made mainly of SiO$_2$ at the equator and larger particles dominated by Mg$_2$SiO$_4$ at mid-latitudes. However, these works also found that the aerosol distribution was much more longitudinally homogeneous, in contrast with observations. This may be due to non-convergence of some of the processes considered in the models.

An important takeaway of the results of 3D models is that the aerosol distributions on the east and west limbs of hot Jupiters are unlikely to be the same, which could be observable via transmission spectroscopy \cite{line2016,vonparis2016,kempton2017}. Using the aerosol microphysics model \texttt{CARMA} combined with temperature profiles extracted from a GCM, \citeA{powell2018} showed that higher temperatures on the east limb promote cloud formation at higher altitudes, making it appear more cloudy than the cooler west limb, which hosts more massive but lower altitude clouds; this shifts abruptly above a critical temperature ($\sim$1700 K), determined by atmospheric circulation, however, when the clouds on the east limb become optically thin in transmission. As such, transmission spectra near this critical temperature may be especially diverse. Detection of patchy aerosols at exoplanet limbs have been difficult in broadband photometry due to a degeneracy between asymmetric limb atmospheres and {uncertainties} in the transit ephemeris, but this may be overcome by taking into account the chromatic variations in how aerosols impact transmission spectra \cite{powell2019}. 

Interpreting possible trends in transmission spectra is complementary to interpreting trends in emission and reflection. While the vertical extent of aerosols is vital in controlling the shape of transmission spectra, it does not strongly affect the emitted and reflected flux \cite{parmentier2016}. As the vertical extent of aerosols is deeply connected to microphysical processes \cite{ackerman2001}, a kinetic model is needed. Using \texttt{CARMA}, \citeA{gao2020} computed the amplitude of the 1.4 $\mu$m water feature of hot Jupiters in transmission as a function of temperature, gravity, and atmospheric metallicity. Their results compare well to the data compiled by \citeA{fu2017}, though there is more scatter in the data (Figure \ref{fig:h2oamp}), which could be due to their usage of a 1D model that ignores east-west limb differences \cite<e.g.>{powell2018}. The computed water feature amplitudes do not vary monotonically with temperature: \texttt{CARMA} predicts the formation of silicate, corundum, and titanium clouds at high temperatures, rapidly reducing the water feature amplitude compared to hotter, cloudless (1D) atmospheres; this is followed by the sinking of these clouds to lower altitudes at lower temperatures leading to an increase in the water feature amplitude; finally, at $\sim$950 K, photochemical hazes form from methane photolysis, reducing the water feature amplitude once more. Importantly, \citeA{gao2020} predicts that optically thick iron and sulfide clouds, including MnS clouds, are difficult to form due to energy barriers associated with nucleation. These results are in contrast to those of \citeA{parmentier2016}, who required silicate clouds to disappear for $T_{eq}$ $<$ 1600 K and for MnS to become the primary aerosol species on the western limb. While a similar transition occurs in the work of \citeA{gao2020}, it is set at a much lower temperature ($\sim$950 K) and is between silicates and methane-derived hazes rather than silicates and MnS clouds. A possible solution is if hazes could form at higher temperatures \cite<e.g.>{lavvas2017} such that it could replace MnS as the main source of aerosol opacity at the limb, while remaining optically thin in emission. An explanation for how silicate clouds disappear is also needed, as sequestration at depth may be difficult \cite{thorngren2019}. In addition, \citeA{gao2020} does not explain the diversity of spectral slopes in the optical. \citeA{ohno2020} offers a possible cause for such slopes by appealing to photochemical hazes. They found that variations in haze formation rates at high altitudes and the rates with which haze particles are mixed downwards naturally lead to a diversity of optical spectral slopes. In particular, they showed that ``super-Rayleigh'' slopes {\cite{pinhas2019,welbanks2019,may2020,alderson2020,chen2021}} are possible when mixing is strong and the haze formation rate is moderate. 

%\citeA{gao2020} makes several simplifying assumptions, chief among which is the reliance on a 1D model to simulate an inherently 3D system. 

%with some studies claiming the possible existence of trends in aerosol distributions with temperature \cite{heng2016,stevenson2016,fu2017,barstow2017,pinhas2019}. 

Modeling efforts for cooler, lower mass exoplanets have focused on understanding why certain benchmark objects, e.g. the mini-Neptune GJ 1214b and the ``super-puffs'' Kepler-51b and d have extremely flat transmission spectra \cite{kreidberg2014,libbyroberts2020}. A slew of studies have attempted to explain GJ 1214b's transmission spectra using 1D models, some relying on KCl and ZnS clouds \cite{gao2018b,ohno2018,ohno2020agg}, some relying on photochemical hazes \cite{kawashima2018,adams2019,kawashima2019a,lavvas2019}, and others relying on both \cite{morley2013,morley2015}. In general, a moderately high ($>$100 $\times$ solar) atmospheric metallicity is needed in addition to aerosol opacity to suppress the amplitude of molecular features to match the data. Studies focusing on clouds have required them to be extremely vertically extended, either due to extremely strong vertical mixing or low sedimentation velocities caused by high porosity. \citeA{charnay2015b} showed using a GCM that cloud particles can be lofted to low pressures by atmospheric circulation to explain the \textit{Hubble} data \cite{kreidberg2014}, but the particle size required would be too small to explain the \textit{Spitzer} data, which extends GJ 1214b's featureless transmission spectrum out to 5 $\mu$m. Studies that explored the impact of hazes frequently faced the same issue: the production of hazes at low pressures leads to small particles that are unable to explain the full spectrum. \citeA{adams2019} was able to explain the full spectrum with aggregate haze particles, though they relied on a parameterization of how the fractal dimension of aggregates scaled with the number of monomers within the aggregate that may not be realistic \cite{ohno2020agg}. \citeA{morley2015} was able to match the \textit{Hubble} data for GJ 1214b with photochemical hazes and predicted that their existence may lead to atmospheric heating, resulting in a temperature inversion that would generate emission features in GJ 1214b's thermal infrared spectrum. In addition, \citeA{morley2015} showed that the haze production rate, parameterized from a photochemical model, peaks at a planet $T_{eq}$ $\sim$ 800 K with falling rates at higher temperatures due to lower methane abundances, and lower temperatures due to decreasing high energy UV photons. This could lead to increasing haze opacity with decreasing $T_{eq}$ for $T_{eq}$ $<$ 1000 K, which may explain the emerging trend in the amplitude of the 1.4 $\mu$m water feature seen in \citeA{crossfield2017}.

The low gravities of super-puffs \cite<e.g.>{masuda2014} led to expectations of large ($>$1000 ppm) amplitude spectral features in transmission, but observations showed a flat spectrum instead \cite{libbyroberts2020,chachan2020}. Generating flat spectra for these objects using aerosols is difficult due to the extremely low pressures ($<$1 $\mu$bar) where they must persist. \citeA{wang2019,gao2020sp} got around this issue by taking into account the outward wind that could exist on super-puffs due to ongoing atmospheric loss, which could entrain aerosol particles and push them to higher altitudes. In particular, \citeA{gao2020sp} showed that such a phenomenon could occur on all young, low mass, temperate ($T_{eq}$ $<$ 1000 K) planets, leading to an increase in the radius of a hazy-covered planet, as seen in transit, by as much as a factor of 2 compared to its clear-sky equivalent. On the other hand, nadir-geometry observations (emission and reflection) may be able to see past some of the aerosol opacity to reveal gas compositions and a smaller radius. 

% 
%The peculiar atmospheric conditions on tidally locked planet strongly shape the horizontal cloud distribution. The first factor is a thermal effect: 

%%This could be moved somewhere else, it's the main conclusion of Parmentier2016. We should discuss that this is in contrast with Gao2020 conclusion that silicate fits them all. The global picture is complicated  ! 

\subsubsection{Directly Imaged Exoplanets}

An important motivation of aerosol models of the current sample of directly imaged exoplanets is to explain why they are redder than brown dwarfs of the same effective temperature, given the general picture of cloud evolution on brown dwarfs outlined in ${\S}$\ref{sec:obs}. \citeA{marley2012} showed that the difference in gravity between field brown dwarfs and directly imaged exoplanets is the likely culprit: the atmospheric mass - and thus opacity - above a given pressure level is higher for a low gravity object than for a high gravity object, leading to the former object having higher temperatures at all pressures than the latter object for the same effective temperature. As such, the cloud base on the lower gravity object would be situated at lower pressures, allowing clouds to persist above the photosphere of the object to a lower effective temperature, leading to redder near-infrared colors due to aerosol opacity. \citeA{charnay2018} reaffirms this result but also shows that the location of the radiative-convective boundary is at lower pressures for lower gravity objects than for higher gravity objects of the same effective temperature, and thus lofting of cloud particles {by convective turbulence} may be more efficient for low gravity objects, further increasing cloud opacity. It is important to note, however, that these studies do not consider the kinetics of cloud formation, {and thus how cloud opacity varies with gravity on directly imaged exoplanets is still uncertain. Furthermore, interpreting spectra of the reddest objects still requires the inclusion of high altitude submicron aerosols and/or highly vertically extended cloud layers  \cite{hiranaka2016,lew2016,kellogg2017,burningham2017,schlawin2017,manjavacas2018,wardduong2020,stone2020}.} 

{In addition to mineral clouds, the proximity of some directly imaged companions to their young, UV-bright host stars coupled with relatively cool stratospheres could permit the formation of photochemical hazes. \citeA{griffith1998} hypothesized that absorbing organic hazes could persist at several tens of bars in the atmosphere of the brown dwarf companion Gl 229 B, which would explain its low flux at red optical wavelengths. Photochemical modeling of directly imaged exoplanets \cite{moses2016,zahnle2016} showed that optically thick hydrocarbon hazes are difficult to form, though there is great uncertainty in the chemical pathways involved. Interestingly, \citeA{zahnle2016} showed that H$_2$S photochemistry could produce elemental sulfur allotropes like S$_8$, which may condense to form optically thick sulfur clouds for planets with effective temperatures $<$700 K. Such clouds would be highly reflective at red-optical and near-infrared wavelengths, but highly absorbing at wavelengths $<$0.4 $\mu$m \cite{gao2017sulfur}. } 

Spatial inhomogeneity and temporal variability of cloud distributions on brown dwarfs and directly imaged exoplanets have been used to explain the variability in rotational light curves of these objects (${\S}$\ref{sec:obs}), but what causes the inhomogeneity is uncertain. \citeA{showman2013} showed using a cloudless GCM that there are likely large-scale upwelling and downwelling regions on these objects that would serve as areas of cloud formation and cloud depletion, respectively. \citeA{tan2017} included a parameterization of silicate condensation and latent heating in their GCM study and found that isolated silicate storms can occur when the condensation level sinks below the radiative-convective boundary due to the onset of moist convection, which could explain the inferred patchiness of clouds and temporal variability of objects at the L-T transition. Variability is also likely impacted by the rotation rate and cloud radiative feedback {\cite{tan2019,tan2020,tan2021}}, such that changes in the coriolis force with latitude could lead to corresponding changes in cloud opacity and patchiness inline with observations of brown dwarfs at different inclinations, where objects viewed equator-on are redder and more variable than objects viewed pole-on \cite{vos2017}.

%The increased redness implies higher aerosol opacity, which would extinct shorter wavelength emitted flux (J band) more than longer wavelength emitted flux (K band), while the increased variability could point to cloud breakup at earlier spectral types than field brown dwarfs. 

%These observations demonstrate that aerosol evolution with effective temperature and age occurs differently between high gravity field brown dwarfs and low gravity objects such as directly imaged companions and young brown dwarfs. 

%vivien.parmentier: Are these your conclusions ? Or is that based on some papers ? Could do with citations in the seconf case.

%vivien.parmentier: That's really interesting but hard to get from the technical discussion above. Could we have a figure showing this ?

%gravity dependence and vertical mixing (DI exoplanets) \cite{skemer2012,marley2012}, position of clouds and how it explains spectroscopic variability 

%On objects that are not tidally locked, such as young directly images planets or Brown dwarfs or on planets too cold to sustain large thermal contrast (e.g. tidally locked temperate planets) the atmospheric transport is more likely to determine the cloud spatial distribution than the thermal structure. 

%\cite{oreshenko2016}

In summary, exoplanet aerosol models have shown that aerosols are intimately linked with the atmospheric thermal structure and vertical mixing. Using models with a wide range of complexity, we have found that the composition of exoplanet aerosols likely transitions between several major compounds with temperature, including silicates, sulfides, and photochemical hazes. In addition, 3D models have shed light on the complexity of aerosol dynamics in both transiting and directly imaged exoplanets. However, all of these models contain important assumptions on how aerosols form and evolve, many of which require laboratory experiments to validate. In the next section, we summarize the laboratory studies that have shed light on the potential complexity and diversity of exoplanet aerosols.

%By warming up the atmosphere and changing the thermal gradient aerosols can create the conditions for their own disappearance leading to potential oscillations in the cloud content of brown dwarfs atmospheres \cite{tan2019}.

\section{Insights from Laboratory Studies}\label{sec:lab}

% New He et al. paper 

The numerous parameterizations made by models in simulating exoplanet aerosols demonstrate the complexity in aerosol formation and evolution in exoplanet atmospheres, complexity that can often only be unveiled by experimental work. The few laboratory exoplanet studies that have been performed thus far {have primarily focused on the formation and composition of hazes, as inspired by similar solar system studies such as investigations of Titan's hazes \cite<see>[for a review]{cable2012}, where haze formation in N$_2$/CH$_4$ gas mixtures at $<$300 K are considered. These works} all involve exposing various gas mixtures in a chamber under vacuum to an energy source, which dissociates and ionizes molecules that can then recombine and grow into larger haze particles. Experiments cover a range of possible atmospheric compositions and temperatures, from those of hot Jupiters to terrestrial planets. Each experimental study is distinct in its choice of temperatures, pressures, gas mixtures, gas flow mechanisms, and irradiation sources and thus drawing larger trends out of their results remains challenging at present. Since the actual atmospheric compositions of exoplanets are currently only loosely constrained, the choice of gas mixtures in many of these experiments is either based on equilibrium model predictions or earlier solar system studies.

\begin{figure}[hbt!]
\centering
\includegraphics[width=0.6\textwidth]{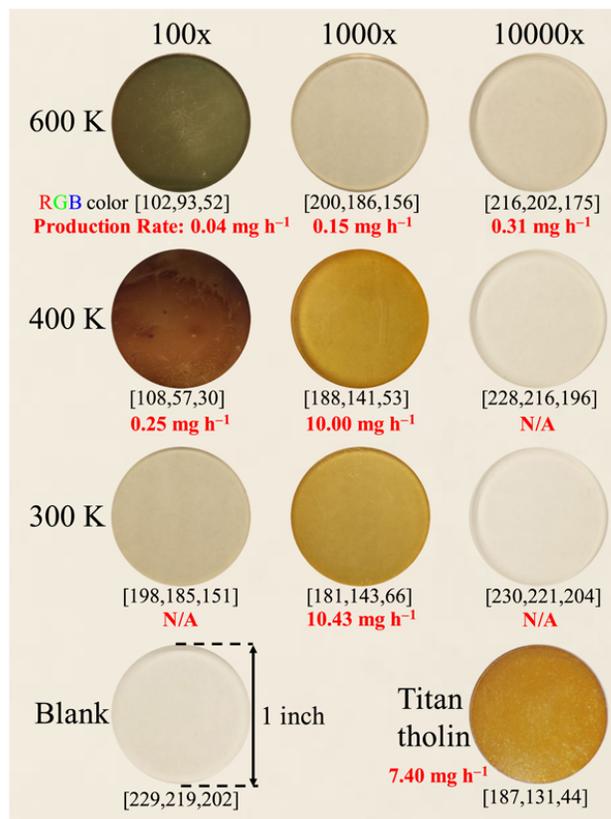}
\caption{Laboratory hazes made from hydrogen-rich, water-rich, and carbon dioxide-rich atmospheres from 300 K to 600 K have a range of colors at visible wavelengths, some unlike those seen in solar system hazes \cite{he2018a}. \copyright~AAS. Reproduced with permission.}
\label{fig:labhazecolor}
\end{figure}

\citeA{fleury2019} measured the photochemical output of a simple atmosphere of H$_2$ with 0.3\% CO between temperatures of 600 K and 1500 K exposed to UV (Ly$\alpha$, 121.6 nm) photons to simulate photochemistry in hot Jupiters. No solid aerosol material was observed for most of their temperature range except at 1473 K and after very long UV exposure times, though contamination by the ambient atmosphere may have influenced their results. {\citeA{horst2018b,he2018b,he2020,he2020bpsj}} conducted a series of experiments targeting hazes in mini-Neptunes and rocky planets with temperatures between 300 and 800 K incorporating gas mixtures dominated by H$_2$, H$_2$O, and CO$_2$, with varying amounts of CH$_4$, CO, NH$_3$, N$_2$, and H$_2$S, as determined by equilibrium chemistry calculations. Both plasma discharge and UV energy sources were used. These experiments showed that increasing H$_2$ tended to decrease aerosol particle production, while the water-dominated atmospheres actually produced more haze than Titan experiments, suggesting that some temperate terrestrial atmospheres may be extremely hazy \cite{horst2018b,he2018b}. The visible appearance of these haze materials are highly diverse, as shown in Figure  \ref{fig:labhazecolor}, hinting at a similar diversity in optical properties and compositions. In addition, the inclusion of sulfur species were found to dramatically increase haze production in terrestrial atmospheres and result in organosulfur haze compositions {\cite{he2020,vuitton2021}}, instead of the elemental sulfur allotropes (e.g. S$_8$) that have been predicted by some photochemical models \cite{hu2014,zahnle2016}. These conclusions are consistent with results from a recent, solar system-focused, experimental study of N$_2$/CH$_4$/H$_2$S gas mixtures \cite{reed2020}. Furthermore, the gas phase compositions resulting from these mini-Neptune and super-Earth experiments include a substantial abundance of organic species \cite{he2019}, which are then incorporated into the solids {\cite{moran2020,vuitton2021}}. \citeA{moran2020} demonstrated that oxygen is readily integrated into the haze particles along with nitrogen and carbon when oxygen-carrying gas species are present, which is consistent with previous studies investigating oxidized solar system hazes \cite{trainer2006,hasenkopf2010,horst2014,ugelow2018}. Finally, preliminary characterization of the results of these exoplanet experiments suggests the production of a plethora of prebiotic species, such as amino acids, sugars, and nucleotide bases \cite{moran2020}.

%These laboratory experiments showing the differing compositions as a result of initial gas mixtures and temperatures imply that the resulting optical properties and influence of photochemical hazes on spectra will be equally diverse. In the visible, the mini-Neptune/super-Earth laboratory hazes are a mixture of colors, for example (see Fig. \ref{fig:labhazecolor}). 

Critically, exoplanet aerosol experiments have demonstrated that methane, long used in the exoplanet literature as an essential component of haze formation \cite<e.g.>{morley2015,kawashima2018,gao2020}, is not always needed to produce substantial amounts of haze {\cite{horst2018b,he2018b,fleury2019,he2020,he2020bpsj}} and that exoplanet hazes likely contain more than just hydrocarbons or by-products of methane photolysis {\cite{moran2020,reed2020,vuitton2021}}. While methane may be an intermediary gas product in some of the experiments that use CO and CO$_2$ as the primarily carbon reservoir \cite<e.g.>{fleury2019}, gas phase results show that it is not part of the chemical pathway in all cases, which instead seem more dependent on CO or CO$_2$ photolysis \cite{he2019,he2020}. Additionally, photochemical models are typically limited to hydrocarbon species containing up to only five carbon atoms \cite<e.g.,>{arney2016,zahnle2016} or even fewer \cite{kawashima2018}, but laboratory work focusing on Titan hazes shows that higher order reactions must be considered to realistically capture aerosol growth \cite{berry2019}. Taken as a whole, laboratory results have clearly shown that the formation of haze in exoplanet atmospheres is not nearly as simple as that assumed in previous and current models \cite<e.g.>{gao2020}, and that a greater appreciation for the chemistry and physics of haze formation at high temperatures is warranted. 

%These results may complicate the idea that a transition exists from silicate clouds to hydrocarbon hazes in light of the possible increase of CH$_4$ below 950 K \cite{gao2020}. However, both the hot Jupiter results \cite{fleury2019} and the mini-Neptune/super-Earth results \cite{moran2020} tentatively suggest that increased temperatures result in refractory haze particles, which could be consistent with hydrocarbon ``soots'' also sometimes used in modeling studies \cite<e.g.,>{morley2013,lavvas2017,gao2020sp}. Many more experiments must be performed to fully validate this hypothesis in conjunction with observations to test the proposed hydrocarbon haze to silicate cloud transition.

%someone please do all the observations and help me do all the experiments to test this; it keeps me up at night

{Measuring the optical properties of exoplanet aerosol materials allow for a direct link to observations of exoplanet atmospheres and facilitates calculations of aerosol opacity in exoplanet atmospheric models. While refractive indices of a variety of cloud compositions exist \cite<e.g.,>{wakeford2015}, these measurements are not necessarily representative of exoplanet atmospheric conditions. Meanwhile, the most frequently used \cite<e.g.>{sudarsky2000,sudarsky2003,howe2012,sing2013,morley2015,wakeford2015,kitzmann2018,kawashima2018,kawashima2019b,adams2019,ohno2020,gao2020sp} set of haze refractive indices in exoplanet investigations have come from the work of \citeA{khare1984}, who measured the optical properties of Titan haze analogs (``tholins''). Other less frequently used optical properties include that of soots \cite{morley2013,lavvas2017,gao2020}, also made predominately of hydrocarbons.}  

%from 250 Angstroms to 1000 microns. Laboratory investigations of Titan's hazes \cite<see>[for a review]{cable2012} primarily considered N$_2$/CH$_4$ gas mixtures at 300 K and cooler. 

%{Optical properties for a wide variety of cloud condensates exist \cite<e.g.,>{wakeford2015}, though these measurements are not necessarily representative of exoplanet atmospheric conditions. However, as these measurements do not yet exist, we leave their discussion to Section \ref{sec:future} and focus here on the existing exoplanet haze studies.}

\citeA{gavilan2017,gavilan2018} conducted spectroscopy and ellipsometry of solid material produced from essentially Titan-like atmospheres at 300 K, with the addition of CO$_2$. They found that the aerosols they made were composed of complex organics, with prominent amide, hydroxyl, and carbonyl groups. In addition, the increased oxidation of the hazes were found to strongly increase their absorptivity in the UV and the mid-IR, particularly between 0.13 and 0.3 $\mu$m and 6 and 10 $\mu$m, as well as blueshift the absorption edge from the visible to the UV, consistent with an early Earth experiment of similar composition \cite{hasenkopf2010}. In contrast, another similar composition early Earth-focused laboratory study, but which also contained molecular oxygen, found no UV absorption from oxidized hazes, though their experimental set-up limited their results to discrete wavelength measurements at 405 and 450 nm \cite{ugelow2018}. These works constitute the only measurements of spectra or refractive indices of exoplanet haze analogs thus far and likely represent only a tiny fraction of the potential diversity of haze optical properties. Solar system studies have shown that gas composition, pressure, temperature, and energy source all impact the spectral response of the resulting haze particles \cite{imanaka2004,brasse2015}. 

%\subsection{Haze Microphysics}\label{sec:labparticles}

\begin{figure}[hbt!]
\centering
\includegraphics[width=0.6 \textwidth]{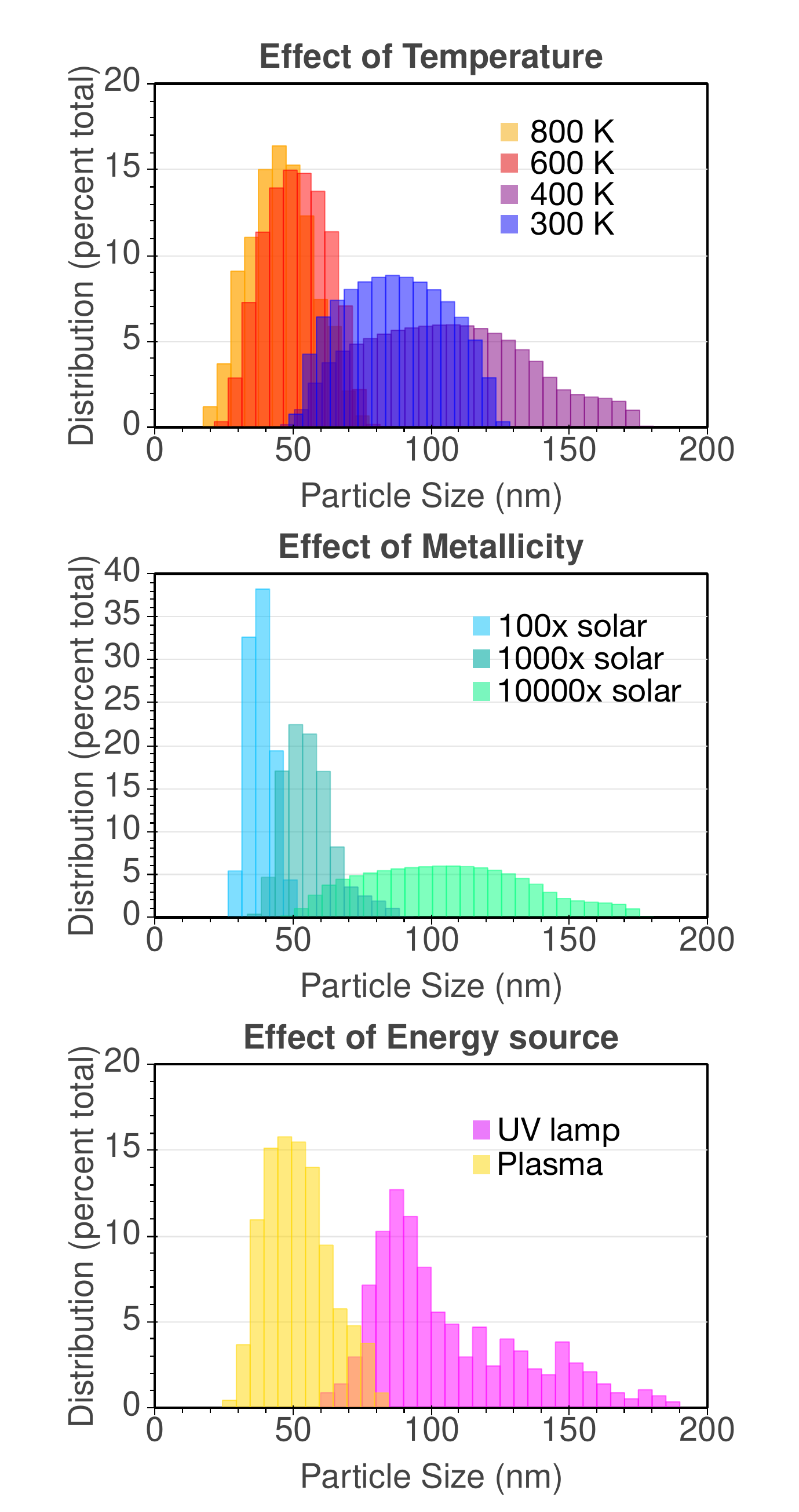}
\caption{Summary of particle size distributions from the laboratory haze experiments of \citeA{he2018a,he2018b,he2020} for 300--800 K {(top)}, 100$\times$--10000$\times$ solar metallicity {(middle)} atmospheres bombarded by UV photons and plasma discharges {(bottom)}.}
\label{fig:labsizes}
\end{figure}

The particle size distribution of aerosols offer a glimpse of the microphysical processes involved in aerosol formation and growth. Size distributions measured by \citeA{he2018a,he2018b,he2020} for the temperate, high metallicity mini-Neptune and super-Earth atmospheres were unimodal and ranged between 20 and 200 nm for the temperatures, initial gas mixtures, and energy sources considered (Figure \ref{fig:labsizes}), which would be able to produce spectral slopes in optical and near-infrared exoplanet transmission spectra. Size distributions were wider for experiments conducted with UV as the energy source than for those conducted with plasma, but the plasma experiments generated more particles. This variance in particle sizes likely results from the difference in energy densities imparted by the UV versus the plasma discharge, but extrapolation to meaningful proxies for diverse stellar types is unclear. High temperatures produced narrower size distributions than cooler temperatures, but the cooler temperatures bore the largest particles. Higher metallicity atmospheres produce both more and larger particles, suggesting that the increased chemical complexity of the atmosphere is able to generate increasingly large, complex molecules. This is further displayed with the addition of sulfur in the form of H$_2$S to the initial gas mixture, which resulted not only in increased particle production \cite{he2020}, but also in larger particle effective densities \cite{reed2020}. Microscopy of the particles showed that not all of them are spherical, and that some particles clump into more aggregate structures, while some form linear chains. Though this is qualitatively consistent with modeling studies that consider aggregate particles \cite{arney2016,adams2019,ohno2020}, the specific growth mechanisms of exoplanet hazes made in the laboratory remains highly uncertain past their initial formation, and the dynamics of haze particles in planetary atmospheres are unlikely to be fully captured by current experimental studies. 

%It should be noted, however, that these laboratory hazes were generated at pressures of a few mbar, and so the pressure-dependence of particle size is uncertain. Moreover, the experiments do not fully capture the dynamics of haze particles in the atmosphere, 

%atmospheric dynamics, as discussed in ${\S}$\ref{sec:mod}, can cause particles to move through the atmosphere, collide, coagulate, and grow, whereupon they may sediment out of the atmosphere or be upwardly mixed. 

%\subsection{Clouds in the Lab}\label{sec:labclouds}

\section{Summary and Future Prospects}\label{sec:pic}

\subsection{An Emerging Picture}

Aerosols are fundamental components of exoplanet atmospheres across a wide range of temperatures, gravities, compositions, and ages. The provenance and composition of aerosols vary with planetary parameters, leading to differences in the planets' emitted flux, geometric albedo, and transmission spectra. By combining state of the art observations with the latest theoretical models and laboratory data, we can summarize our current understanding of the nature of exoplanet aerosols.

On tidally locked, transiting exoplanets {with hydrogen/helium-dominated atmospheres}, the longitudinal variation in instellation coupled with atmospheric circulation result in daysides mostly devoid of aerosols for equilibrium temperatures $\geq$1000 K, while the nightside and western limb possess optically thick aerosol layers up to $\sim$2500 K. This spatial inhomogeneity likely generates the observed low dayside albedoes and nightside emission fluxes, as well as the westward shifted brightness maxima in optical phase curves. Clouds of silicates and other oxide minerals likely dominate the total aerosol opacity at high temperatures ($\geq$1500 K), while other types of aerosols, such as sulfide clouds and photochemical hazes, dominate at lower temperatures, causing the observed variations in the spectral slope and the amplitude of molecular features in transmission spectra. {Exoplanet photochemical hazes are likely to possess diverse compositions, incorporating atomic species like carbon, oxygen, hydrogen, nitrogen, and sulfur}. Clouds on the directly imaged exoplanets discovered to date should be similar in composition to their transiting cousins, though the evolution of these clouds with planetary parameters should be more akin to that on brown dwarfs. A major difference is that the lower gravity of directly imaged exoplanets, as compared to brown dwarfs, leads to the persistence of clouds above the photosphere to lower effective temperatures. The sinking and breaking up of clouds likely trigger the L-T transition in brown dwarfs and directly imaged exoplanets and cause the observed temporal variability in emission. 

%Global picture: 
%\begin{itemize}
%\item Daysides are cloud-free : we know that because emission spectra don't need any cloud (WASP-43b from Kataria, HD209 from Line et al. and all the UHJ) and albedos are small (even HD189 can be fitted without clouds, Kepler didn't find ANY planet with a really large albedo). 
%\item Limbs are cloudy, cloud coverage (determine by temperature variation along the limb and particularly east/west dichotomy), cloud vertical extent or cloud variation accross the limb depth are likely responsible for the variety of behaviour we see. 
%\item Nightside are cloudy (Beatty cold nightside temp. we can cite my draft paper, hopefully on ArXiV before the review is accepted :D )
%\item west part of dayside have a crescent of clouds, works well with a cloudy west limb and cloud free east limb. 
%\end{itemize}

%If this is the case, then the brightness temperature as a function of longitude and latitude on a planet would be $\sim$1100 K anywhere the cloud is optically thick throughout the nightside, resulting in a brightness temperature map that is ``flat'' on the nightside \cite{beatty2019}. This provides a solution to a long standing problem where the inversion of observed phase curves using sinusoidal temperature maps resulted in negative temperatures on the nightsides  \cite{beatty2019}. 

\subsection{Outstanding Questions}

While we are now able to construct a coherent picture of the formation and distribution of aerosols in exoplanet atmospheres, there are still many holes in our understanding. Below, we list a number of outstanding questions that will require detailed study in the next decade and beyond: 

 \begin{enumerate}
 \item What are the compositions of exoplanet aerosols? Are they mixtures or mostly pure particles? 
 \item {How porous are exoplanet aerosol particles? Are they dense or fluffy aggregates?}
 \item How do clouds initially form? What is the condensation sequence in exoplanet atmospheres? 
 \item What are the major parent molecules and chemical formation pathways of exoplanet photochemical hazes?
 \item What {physical processes lead to the cloud evolution hypothesized} at the brown dwarf L-T transition and how do these change for directly imaged exoplanets?  
 \item What level of model complexity is necessary to capture the aerosol processes inferred from current and future observations to high fidelity? {How should exoplanet aerosols be parameterized in retrievals? }
 \item How do aerosols respond to variations in planetary, atmospheric, and host star properties? {Conversely, how do aerosols affect the composition, thermal structure, and dynamics of exoplanet atmospheres?}
 \end{enumerate}
 
\subsection{Future Observations}

\begin{figure}[hbt!]
\centering
\includegraphics[width=0.6\textwidth]{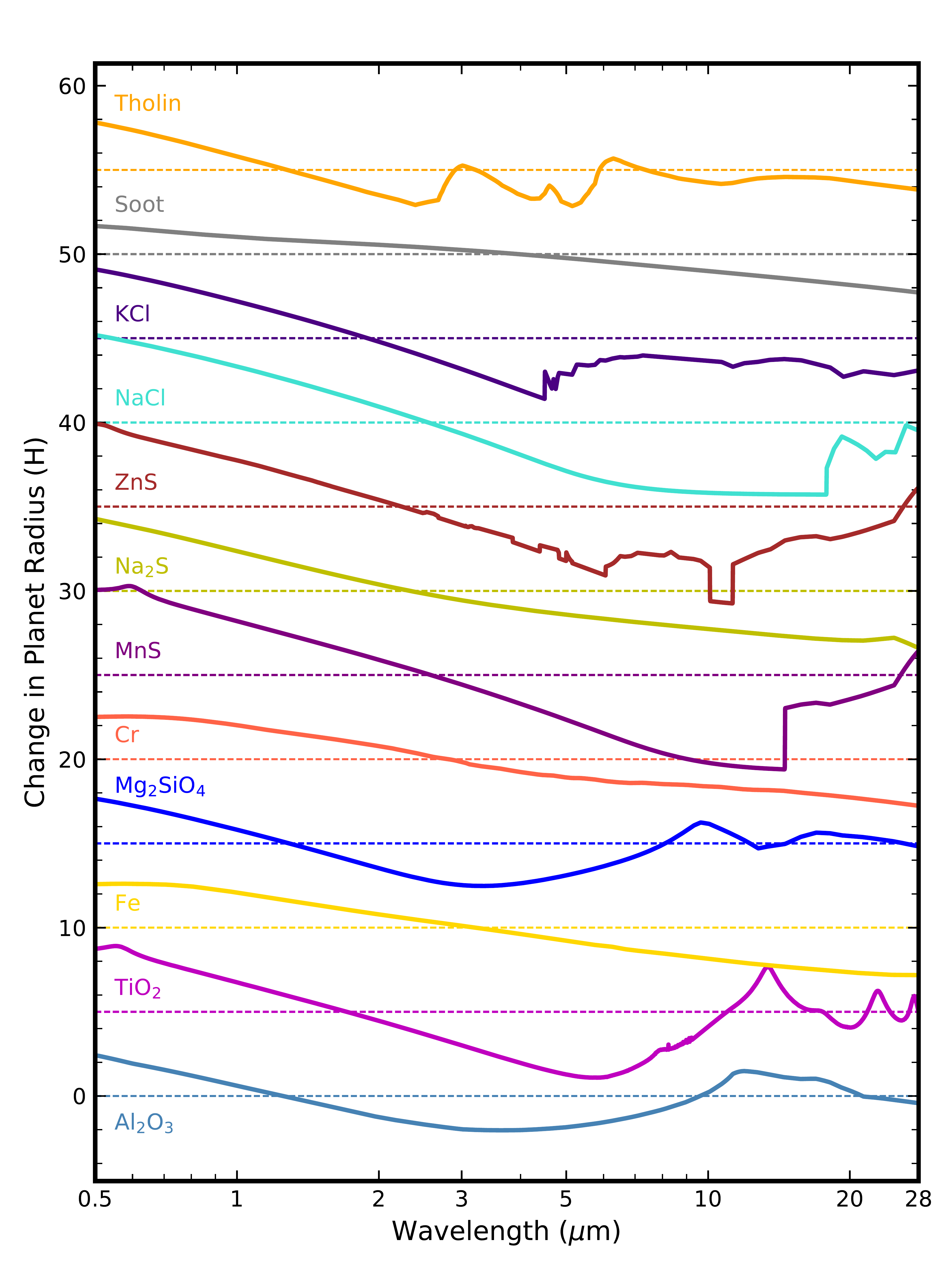}
\caption{Aerosol transmission spectra for a variety of proposed cloud and haze species computed assuming monodisperse 0.1 $\mu$m particles distributed with a constant mass mixing ratio profile in the atmosphere. The spectra are offset for clarity, normalised to the mean transit depth, and shown in planetary scale heights. Optical constants for tholins are taken from \citeA{khare1984}; those for soots are from \citeA{lavvas2017}; those for KCl, ZnS, Na$_2$S, MnS, and Cr are from \citeA{morley2012}; those for NaCl are from \citeA{eldridge1985,querry1987}; those for Mg$_2$SiO$_4$, Fe, and Al$_2$O$_3$ are from \citeA{wakeford2015}; and those for TiO$_2$ are from \citeA{posch2003,zeidler2011}. {The computed aerosol transmission spectra can be accessed from \citeA{gao2021zenodo}}. }
\label{fig:cloudspec}
\end{figure}

A major goal of future observational investigations of exoplanet aerosols should be to unveil their compositions, which are currently unknown due to the lack of any specific spectral features in current observations. {Spectroscopy in the near-to-mid infrared (2-12 $\mu$m)} with the \textit{James Webb Space Telescope} (\textit{JWST}) will be a critical next step to explore the composition of aerosols and their role in a 3D atmosphere. Many of the proposed aerosol species possess spectral features of their own, which are best measured in the mid-infrared and correspond to the vibrational mode between the dominant atoms in the material (Figure \ref{fig:cloudspec}). Silicates such as enstatite (MgSiO$_3$) and forsterite (Mg$_2$SiO$_4$) have vibrational mode absorption dominated by the Si-O bond which produces prominent absorption at $\sim$10 $\mu$m. These vibrational modes have been measured in an array of astrophysical contexts, including in the atmospheres of brown dwarfs \cite{cushing2006,looper2008}. \citeA{wakeford2015} showed using Mie calculations and the optical properties of various cloud forming species that cloud vibrational-mode absorption features could reach observable amplitudes in exoplanet transmission spectra, though the amplitude is a strong function of the mean particle size and the width and shape of the size distribution, with larger particles and wider size distributions leading to smaller amplitudes. Follow up works investigated additional cloud species {\cite{wakeford2017refractory,kitzmann2018}}, the impact of different sized particle populations \cite{mai2019}, differences in cloud opacity at optical wavelengths \cite{pinhas2017}, {and the impact of taking into account cloud microphysics \cite{ormel2019,gao2020}.}

{The observability of cloud spectral features will also depend on whether the cloud particles are pure, as predicted by equilibrium models, or mixtures, as predicted by kinetic cloud models. \citeA{helling2006obs} showed that consideration of kinetic cloud formation and mixed grains could result in the condensation of cloud species that are suppressed in equilibrium models, such as SiO$_2$, which exhibit mid-infrared absorption features different from those of enstatite and forsterite, the major silicate clouds predicted by equilibrium models. As such, the extent to which exoplanet cloud formation follows equilibrium or kinetic models may be testable using observations. However, it is not yet known how we can go a step further and, in the event that exoplanet clouds are better reproduced using kinetic models, use observations to differentiate between well-mixed cloud particles, like those modeled in \texttt{DRIFT}, and layered cloud particles, like those modeled in \texttt{CARMA}. In addition, the porosity of the cloud particles will also impact the spectral features, with more aggregate-like particles exhibiting stronger absorption \cite{samra2020}.} 

%will be more of a challenge, as  degree to which these two approaches differ in how they impact the rest of the atmosphere has not been studied.

Photochemical hazes may {also} exhibit spectral features in the {near and} mid-infrared \cite<Figure \ref{fig:cloudspec}; also see e.g.,>{wakeford2015,kawashima2018,gao2020sp}, which could shed light on their complex compositions by revealing the types of bonds that they contain. {Many organic polymers that contain mixtures of carbon, oxygen, nitrogen, and hydrogen possess absorption features at wavelengths $\sim$3 $\mu$m and $>$6 $\mu$m \cite<e.g.>{wang1998,laskina2014} corresponding to the vibrational modes of the various single and double bonds between C, O, N, and H in various functional groups. However,} additional laboratory work is needed to measure the optical constants of exoplanet haze analogues before we can predict the amplitude, {width, and exact} locations of these features {and interpret future exoplanet haze observations.}

%The observability of cloud spectral features will also depend on the strength of the adjacent gaseous opacities as compared to the cloud opacities \cite{gao2020}, as well as whether the particles are pure or mixtures. {\citeA{helling2006obs} showed that consideration of kinetic cloud formation and mixed grains could yield cloud species that are suppressed in equilibrium models, such as SiO$_2$, which exhibit mid-infrared absorption features. As such, whether cloud formation follows equilibrium or kinetic models may be testable using observations.   }

In addition to composition, \textit{JWST}'s ability to continuously observe targets at high time cadence, as opposed to \textit{Hubble}'s staccato way of observing brought on by its orbit around the Earth, will allow us to probe the east and west limbs separately. This will be critical for deciphering the 3D distribution of aerosols and how asymmetric limbs impact transmission spectra \cite{fortney2010,line2016,vonparis2016,kempton2017,powell2018,powell2019}.

%ALSO MEASURE CLOUD OPTICAL PROPERTIES AT APPROPRIATE T AND P 

%Whether or not cloud absorption features are actually observable will depend on the individual and combined strengths of aerosol opacities and gaseous opacities at wavelengths where vibrational mode features peak \cite{gao2020}. JWST will be the first opportunity to measure these vibrational modes for exoplanet atmospheres.

%A number of works have followed Building on this work the optical properties of  clouds have been used in a number of studies to further examine current measurements and make predictions for \textit{JWST} \cite<e.g.>{pinhas2017,kitzmann2018,mai2019}. 

%Additionally, the clouds themselves are not unbiased absorbers, the minerals they are composed of have their own unique fingerprints just like the gaseous species. 

%the disentangling of east/west limb asymmetrical limb effects in transmission, leading to probes of how aeroso  meaning the east/west/morning/night limbs of the terminator can be disentangled. While \textit{Spitzer} also had this capability, \textit{Webb}'s spectroscopic capabilities combined with the full transit coverage can be leveraged to disentangle 3D effects on the transmission spectrum of exoplanet atmospheres \cite{fortney2010,line2016,kempton2017,powell2018,powell2019}. 

% Hannah still working on more more more more more....

An alternative strategy for probing the composition of clouds is to look for the existence or absence of gas species that have been hypothesized to condense. In particular, as several groups of gasses are associated with clouds that condense at similar temperatures on hot Jupiters (e.g. TiO/VO, aluminum, and calcium at the highest temperatures, iron, magnesium, silicon, chromium, and manganese at moderate temperatures, and potassium and sodium at lower temperatures, see Figure \ref{fig:cloudcomp}), measuring the absolute abundances of these gases and their ratios as a function of planetary temperature and gravity could help constrain the condensation sequence in exoplanet atmospheres \cite{lothringer2020}. However, while many species have been detected for ultra hot Jupiters \cite<e.g.>{fossati2010,haswell2012,hoeijmakers2018,yan2019,vonessen2019,sing2019,benyami2020,cabot2020,nugroho2020}, suggesting largely cloud-free atmospheres, efforts at lower temperatures have yielded mixed results due to controversial detections that are difficult to replicate {\cite<e.g.>{vidalmadjar2013,cubillos2020,sedaghati2017,espinoza2019,chen2018,seidel2020,sing2015,gibson2017,gibson2019,mcgruder2020}} and aerosol opacity at optical wavelengths that reduce the amplitudes of atomic and molecular absorption features \cite{charbonneau2002,pont2008,heng2016,sing2016}. 

Future high spectral resolution observations in the optical and near-ultraviolet by groundbased extremely large telescopes and space-based telescopes will be essential for accurate measurements of heavy element abundances in the upper atmospheres of exoplanets. The potential for constraining cloud and dynamical processes on hot Jupiters with high spectral resolution observations was shown by \citeA{ehrenreich2020}, who recently detected the blue-shifted spectral signature of neutral iron on the eastern/dusk limb of the hot Jupiter WASP-76b but not on the western/dawn limb. This is suggestive of iron condensation on the nightside of the planet after it was transported there by eastward winds. In addition, high spectral resolution data can reveal the vertical distribution of aerosol layers, particularly for planets that exhibit {low-resolution} flat optical or near-infrared transmission spectra, by revealing the cores of spectral lines that extend above the aerosols {\cite{pino2018,hood2020,gandhi2020}; this was recently attempted for some hot Jupiters \cite{sanchezlopez2020,allart2020}.}

Future reflected light observations by groundbased extremely large telescopes and space-based telescopes will also allow for probes of exoplanet aerosols. \citeA{morley2015,charnay2015b} showed that mini-Neptunes like GJ 1214b that possess flat near-infrared transmission spectra may exhibit a variety of optical geometric albedo spectra that are diagnostic of aerosol compositions and particle sizes. \citeA{morley2014,macdonald2018,hu2019} argued that water clouds on cooler giant exoplanets can boost their albedo such that the water and methane absorption bands in the red-optical become much more prominent, aiding retrievals of molecular abundances. {However, degeneracies between retrieved cloud properties and molecular abundances could arise unless the planet could be observed at multiple orbital phases \cite{carriongonzalez2020,damiano2020,damiano2020reflect}.} {Sulfur clouds sourced from H$_2$S photochemistry can also boost planets' red-optical albedoes, though their blue-optical and near-UV albedoes would be much lower \cite<$\sim$0.1;>{gao2017sulfur}.} \citeA{lacy2020} revealed that silicate clouds will greatly affect the optical spectrum of current directly imaged exoplanets by increasing their brightness and muting molecular and atomic absorption features. \citeA{mayorga2019} found that optical phases curve amplitudes in the \textit{TESS} bandpass tend to be low ($<$10 ppm) for a variety of cloud species, but that amplitudes increase towards bluer wavelengths. In addition, measurements of the polarization of reflected light offer unique constraints on the composition, size, shape, and spatial distribution of aerosol particles \cite{seager2000curve,karalidi2013,kopparla2016}, though claims of detections so far have been controversial \cite<e.g.>{berdyugina2011,wiktorowicz2015,bott2016}. Polarization {of the thermal emission of brown dwarfs and directly imaged exoplanets can also constrain aerosol properties and distributions in their atmospheres} \cite<e.g.>{sengupta2001,marley2011,stolker2017,sanghavi2018,millarblanchaer2020}. 

%particle size constraints?
%- Vivian has some stuff from emission.

Further constraints on aerosol particle sizes and vertical distribution, important for shedding light on a myriad of microphysical and dynamical processes in the atmosphere, can be gleamed from current and future transmission and emission spectroscopy. Small particle sizes ($\leq$0.1 $\mu$m) have already been estimated from the spectral slopes in optical transmission spectra \cite<e.g.>{lecavelierdesetangs2008,wakeford2015,wakeford2017refractory,wong2020}, while extensions to the \textit{Spitzer} bandpasses have yielded even tighter size constraints \cite{benneke2019}. Observations by \textit{JWST} towards the mid-infrared will enhance these efforts by probing the decrease in opacity of larger particles, while the continuous wavelength coverage will allow for more precise measurements of the shape of the aerosol continuum, which contains information about the vertical distribution of aerosol particle sizes \cite{mai2019}. Meanwhile, tracking the wavelength-dependent light curve variability amplitude of L dwarfs has allowed for the measurement of aerosol particle sizes \cite{lew2016,schlawin2017}, since aerosol opacity impacts shorter wavelengths more than longer wavelengths. Future spectroscopic light curve surveys of L-type objects may be able to constrain cloud particle sizes as a function of effective temperature and gravity.

Finally, detections of weather/temporal variability in exoplanet atmospheres impose powerful constraints on dynamical processes therein, including how aerosol microphysics is coupled to the atmospheric circulation pattern. Future long time baseline, high precision observations will shed light on exoplanet weather in the same way rotational light curve measurements have informed our understanding of weather on L and T type objects. Brown dwarf surveys are already delving into cooler temperatures with measurements of variability on Y dwarfs \cite{rajan2015,esplin2016,cushing2016,leggett2016} that could be probing patchy sulfide and chloride clouds and the onset of water clouds, thus bridging the gap between brown dwarfs and the giant planets in our own solar system. GCMs have predicted hot Jupiter atmospheric variability due to propagating waves and instabilities on spatial scales of thousands of kilometers to global scales {\cite{dobbsdixon2010,parmentier2013,lines2018,komacek2020}}, which could lead to temporal variations in the spatial distribution of aerosols. Brightness variability has also been detected on hot Jupiters \cite{armstrong2016,jackson2019}, with interpretations ranging from clouds reacting to changing winds and temperatures to the coupling of atmospheric circulation with the planets' magnetic field \cite{rogers2014,rogers2017}. 

%hot Jupiter variability (Brian Jackson+ armstrong but also mention this can "just" be MHD moving the actual hot spot east and west of the substellar point and all kepler photons are thermal emission. Clouds could enhance that though (rogers et al. there's two paper). ) Komacek theory paper,  parmentier2013

\subsection{Future Modeling Efforts and Laboratory Studies} \label{sec:future}

The anticipated future exoplanet observations will necessitate the development of more sophisticated aerosol models and more detailed laboratory measurements. 3D models of exoplanet atmospheres coupled with kinetic models of aerosol microphysics that include cloud radiative and latent heat feedback will likely be required to fully understand these new observations due to the wavelength- and spatial-dependence of aerosol opacity. In particular, general, flexible models with the ability to simulate multiple particle size modes and multiple aerosol species (e.g., both clouds and hazes) for various background atmospheric compositions and thermal structures will need to be developed to interpret new data from hundreds, perhaps thousands of exoplanets. At the same time, non-hydrostatic atmospheric models capable of resolving and studying moist convection in H/He atmospheres \cite<e.g.>{freytag2010,li2019,ge2020} will allow for more rigorous investigations of cloud particle and condensate vapor transport, cloud formation, and cloud spatial inhomogeneity and temporal variability on giant exoplanets and brown dwarfs. Advances in {computational efficiency} will be essential for all of these innovations in modeling.

{At the same time, intercomparisons between existing models are needed for understanding best modeling practices, evaluating model consistency, and placing interpretations of observations by different models on a more equal footing. As discussed in ${\S}$\ref{sec:modeldiscussion}, comparisons of different treatments of aerosols in retrieval codes have been recently undertaken, showing a sensitivity of aerosol distribution parameters to the modeling assumptions \cite{mai2019,barstow2020cloud}. \citeA{helling2008a} compared several more complex aerosol models that were (at the time) used mostly for brown dwarf atmospheres, including \texttt{DRIFT} and the \citeA{ackerman2001} model, among others. They found that different models predicted different vertical and particle size distributions of clouds, leading to variations in the predicted brown dwarf near-infrared fluxes of several tens of \%. Given the recent proliferation of complex aerosol microphysics models (Table \ref{tab:micromodels}), an update to \citeA{helling2008a} is warranted. }

{For 3D aerosol models, \citeA{lines2019} juxtaposed the advection of cloud distributions computed from \texttt{DRIFT} by a GCM with stationary cloudy atmospheric columns computed by the \citeA{ackerman2001} model coupled to the same GCM. They found that the predominance of small, high altitude, scattering particles in \texttt{DRIFT} led to a decrease in global temperatures when compared to the larger, lower altitude particles predicted by the \citeA{ackerman2001} model. \citeA{showman2020} compared several different aerosol treatments in GCMs, including passive tracers and microphysical models, and showed that all models found latitudinal variations in aerosol distributions, but whether the aerosol abundance increased or decreased with latitude is model dependent.  }

{In addition to comparisons of models of similar complexity, lessons learned from complex aerosol models should be incorporated into simpler models used in retrievals and radiative-convective equilibrium models so that they can be made more physical and predictive, while allowing for connections between more subtle microphysical processes and exoplanet observations. \citeA{gao2018a} attempted to place the sedimentation efficiency parameter of the \citeA{ackerman2001} model into the context of microphysical processes by comparing it to \texttt{CARMA} for different planetary and aerosol parameters, but a more systematic and extensive effort is needed. }

The measurement of a variety of material properties and aerosol processes through laboratory experiments is critical for interpreting future observations and building more physical models {\cite{fortney2016WP}}. Many of the optical constants currently in use for exoplanet aerosols are decades old and not measured {for the composition and temperature regimes of} exoplanet atmospheres \cite{morley2012,wakeford2015,kitzmann2018}, and thus updates are needed. This issue is exacerbated in the case of exoplanet hazes, which should have a diverse composition (${\S}$\ref{sec:lab}). Saturation vapor pressures of certain exoplanet condensates must also be measured, as many are currently estimated from equilibrium chemistry models alone \cite<e.g.,>{lodders2002tiv,visscher2010}. The same is true for the surface energies of cloud species, which control their nucleation and condensation rates \cite{gao2018a}. In general, condensation in high temperature atmospheres has been rarely explored in the laboratory, and yet knowing which clouds predicted from equilibrium chemistry actually form, how they form, and how different condensates interact with each other \cite<e.g.>{helling2006} are fundamental problems that can only be solved through laboratory experiments. These efforts would also shed light on whether exoplanet cloud particles are crystalline, amorphous, or mixtures of both, whether they are mostly spherical or irregularly shaped \cite{ohno2020agg,samra2020}, and whether they are liquid or solid, all of which affect their optical properties and how they couple to the rest of the atmosphere. In addition, laboratory studies have yet to address the interplay of exoplanet clouds and hazes together as an interconnected set of processes, as seen in the Solar System \cite<e.g.,>{andreae2008,lavvas2011,wong2017}. Modeling studies have suggested that photochemical hazes may in some cases act as cloud condensation nuclei \cite{gao2018b}. Preliminary measurements on the solubility of exoplanet haze analogues suggest that some may in fact enable cloud formation of polar condensates \cite{moran2020}, but verification and generalization of such results are needed. Finally, laboratory efforts will be required to elucidate the photochemical pathways that lead from parent molecules to haze particles for different host star spectra and atmospheric composition. Such an effort will be essential for photochemical modeling of the entire haze formation process.

%smoran14: may want to add more about the photochemical pathway part, both as related to stellar spectrum impacting photochemistry and also reconciling how far we need to get the pathway understood in the lab to connect it to photochemical models and vice versa

%and its extension to the many potential exoplanet cloud species, is currently outstanding.

%the number and composition of which can greatly impact the extent and thickness of a cloud deck \cite{ohno2018}. Whether or not hazes, and other materials such as dust or meteoritic smoke, aid in or suppress exoplanet cloud formation is yet another necessary parameter to determine experimentally. 

%Verifying and extending these measurements to the temperature-pressure regimes relevant to exoplanet atmospheres can help elucidate the exact shape these clouds will take beyond assumptions of spherical particles treatable by Mie theory, such as their crystalline structure if they form as solids or their potentially unique scattering behavior if they form instead as liquid drops, which modelling suggests would substantially impact observations \cite{samra2020}. 

%Additionally, measurements must be made to verify the theoretical vaporization and condensation temperatures of potential refractory clouds. 

%The properties of the kinds of clouds expected from theory (See, for example, Fig \ref{fig:cloudcomp} above) have been mostly inferred from laboratory measurements of these materials at Earth conditions \cite<e.g.,>{wakeford2015,kitzmann2018}. 

%yeah it's weird that I cite myself last, but I swear I didn't mean to. oops.

\subsection{Towards Rocky Worlds}\label{sec:rocky}

The atmospheres of rocky exoplanets are undoubtedly more diverse than the H/He-dominated atmospheres we have explored thus far, suggesting similarly diverse aerosol compositions. Importantly, aerosols in rocky exoplanets serve an additional role in their atmospheres as compared to gas giants: a major control on the surface climate and thus habitability. No molecular features have been robustly observed in rocky exoplanet atmospheres thus far {\cite{dewit2016,southworth2017,dewit2018,ducrot2018,zhang2018zb,diamondlowe2018,wakeford2019trappist1,burdanov2019,diamondlowe2020,diamondlowe2020lhs3448b}}, which could be caused by aerosols, a high mean molecular weight atmosphere, a combination thereof \cite{moran2018}, or a lack of an atmosphere altogether \cite{kreidberg2019}. We are therefore left with theoretical predictions about the aerosols that may be present in these atmospheres and their impacts. 

%DUST

The hottest rocky exoplanets, with $T_{eq}$ $\geq$ 1000 K, are likely to have either no atmosphere or thin atmospheres in equilibrium with a molten surface. These atmospheres would be made of metal oxides like SiO, atomic and molecular oxygen, and atomic magnesium, sodium, iron, and other refractory elements, with abundance ratios dependent on the surface composition \cite{miguel2011,ito2015,kite2016,herbort2020}. Under these conditions, clouds of oxidized minerals (silicates, corundum, perovskite) and alkali salts, much like those proposed for hot Jupiter atmospheres, are likely to form \cite{schaefer2009,schaefer2012,mahapatra2017}.

Rocky exoplanets with temperatures more similar to those in the Solar System may possess more familiar aerosols: clouds of water, sulfuric acid, and CO$_2$ in temperate, oxidizing atmospheres, clouds of hydrocarbons and nitriles, along with organic hazes in cool, reducing atmospheres, and dust elevated into the atmosphere by winds in a variety of atmospheres. Temperate, reducing atmospheres may also host organic hazes {\cite<e.g.>{pavlov2001,trainer2004,wolf2010,arney2016,he2018b,he2020,vuitton2021}}, as well as sulfur (S$_8$) clouds \cite{hu2013}. Many of these aerosols are highly reflective at optical wavelengths, facilitating future reflected light observations. In particular, the detection of bright sulfuric acid clouds may point to active volcanism and the lack of significant oceans on the surface of temperate rocky exoplanets \cite{misra2015,loftus2019}.

Of particular interest are rocky exoplanets orbiting near the habitable zone of M dwarfs, as they are easier to characterize than those around Sun-like stars due to the more favorable planet-to-star radius ratio and short orbital periods. In regards to possible aerosols in their atmospheres, several differences between these planets and their solar cousins need to be considered: (1) they are most likely tidally locked to their host stars \cite{kasting1993}, (2) they experience a prolonged period of high instellation during their host stars' pre-main sequence that could desiccate their upper mantles \cite{luger2015}, and (3) they are subjected to higher fluxes of high energy UV radiation and particle bombardment \cite{france2013}. A consequence of (1) is that a planet with a water ocean could exhibit vigorous convection at the subsolar point, leading to the formation of optically thick water clouds, with its high albedo acting as a stabilizing feedback against increasing instellation \cite{joshi2003,yang2013,way2018}. Such cloud patterns are predicted to be sensitive to planet rotation rate \cite{yang2014,komacek2019water} and lead to significant muting of molecular features in transmission \cite{suissa2020,komacek2020water}. On the other hand, planets dried out from (2) could have clear or dusty atmospheres devoid of water clouds and sulfuric acid clouds \cite{lincowski2018,lustigyaeger2019}. Photochemical haze formation due to (3) depends on how different links in the reaction web that leads from simple parent molecules to haze particles react to the spectral energy distributions of M dwarfs. For example, \citeA{arney2018} showed that haze formation in temperate, anoxic atmospheres is more efficient for an M dwarf host star {compared to a Sun-like host star}, particularly in the presence of organic sulfur compounds. 

{\subsection{Exoplanet--Solar System Synergies}}

{Lessons learned from studying aerosols in the atmospheres of solar system worlds have been and will continue to be vital for our understanding of exoplanet aerosols. As shown in this review, the theoretical frameworks used to understand exoplanet aerosols are heavily influenced by what we know of aerosols in the Solar System, and in fact several exoplanet aerosol models are derived directly from models of Earth and solar system aerosols (e.g. \texttt{CARMA} and the \citeA{ackerman2001} model). Laboratory investigations of exoplanet aerosols are even more intimately linked to efforts to experimentally characterize aerosols closer to home, with many deriving from investigations of Titan's haze.} 

{As the quality of exoplanet observations improve in the coming decades, it would be beneficial for analyses of these data to draw inspiration from efforts to analyze solar system data. For example, \citeA{irwin2015} retrieved the imaginary refractive index of the aerosols in Uranus's atmosphere from reflected light spectra in the near-infrared instead of assuming any particular aerosol composition; this strategy may become necessary for future reflected light and thermal emission observations of exoplanets and brown dwarfs \cite<e.g.>{taylor2020ssa}. In addition, several studies have focused on treating observations of solar system worlds as analogues of future exoplanet data. \citeA{mayorga2016,dyudina2016} investigated how the brightness and color of Jupiter and Saturn varied with phase and found significant deviations from a Lambertian model, indicating complex vertical distributions of aerosol particles, with implications for directly imaging exoplanets in reflected light. \citeA{simon2016,ge2019} analyzed rotational light curves of Neptune and Jupiter, respectively, as analogies of brown dwarf light curves, and discovered that light curve variability is controlled largely by discrete features like the Great Red Spot, and that the shape of the light curve depends strongly on the heterogeneous cloud cover and gas opacity. These efforts not only provide cautionary tales of how complicated planetary atmospheres can be, but they also provide benchmarks to which exoplanet aerosol models can be compared. \citeA{karalidi2015}, for example, applied their mapping code to both Jupiter and brown dwarfs and were able to retrieve several major atmospheric features on the former object. Similarly, \citeA{lupu2016} applied a multi-aerosol-layer retrieval code to reflected light observations of Jupiter and Saturn in preparation for future observations of wide-orbit giant exoplanets, and were able to retrieve methane mixing ratios and cloud single scattering albedos consistent with the observed values for those two planets. As we increasingly focus on cooler targets like Y dwarfs and temperate exoplanets with temperatures and atmospheric compositions approaching that of our own giant planets \cite<e.g.>{morley2018,dalba2019,benneke2019k218b,vanderburg2020}, forging connections with solar system science will become ever more important. }

{As with giant planets, much of our predictions of rocky exoplanet aerosols are based on the examples in our solar system (see ${\S}$\ref{sec:rocky}), including sulfuric acid and water clouds on potential Venus-like and Earth-like worlds, respectively \cite{yang2014,lincowski2018}. Organic hazes inspired by aerosols that may have existed in the Archaean on Earth have also been invoked for exoplanets \cite{arney2017}. Understanding how these aerosols impact surface temperatures and ionizing radiation fluxes are vital for predictions of habitability \cite{yang2013,parisi2004}. Recent discoveries of nearby worlds with inferred temperatures much cooler than that of Earth \cite{ribas2018,damasso2020} have also placed a spotlight on exoplanets that may be similar to the icy satellites and Kuiper Belt objects in the Solar System that are enshrouded by organic hazes, such as Titan, Triton, and Pluto. Several studies have investigated the surface temperature \cite{gilliam2011} and transmission spectra and geometric albedo \cite{robinson2014,checlair2016} of Titan-like exoplanets, all of which are strongly dependent on the haze properties. In addition, \citeA{lora2018} used a photochemical model to show that haze production on Titan-like exoplanets for host stars of various stellar types may be similar, though the photochemical pathways require further laboratory studies. }

%Meanwhile, haze formation on Titan-like exoplanets may be minimally impacted in terms of production rate, but may have different compositions and optical properties due to the differences in relative importance of certain reactions with an M dwarf host star versus the Sun \cite{lora2018}; such variations in haze properties will lead to a diversity of surface temperatures \cite{gilliam2011} and observables \cite{robinson2014,checlair2016,fauchez2019}.  

As we have shown in this review, the study of aerosols in exoplanet atmospheres touches upon nearly every aspect of exoplanet science and every method that we use to learn about exoplanets. In the years to come, the study of exoplanet aerosols must continue to advance on two major fronts: (1) understanding the composition and spatial and size distributions of exoplanet aerosols to better constrain their influence on exoplanet atmospheres and (2) leveraging the impact of aerosols on observations so that we can constrain the gaseous composition and thermal structure of the atmosphere. These two goals are intertwined since the aerosol composition and distribution depend strongly on the overall atmospheric composition and thermal structure, while constraining these two attributes through observations requires a more detailed picture of the nature of aerosols. Accomplishing these goals will demand greater synergy between observations, modeling, and laboratory work, as well as between the exoplanet, solar system, and Earth sciences. Ultimately, deciphering the aerosol puzzle will get us one step closer to understanding the atmospheres and origins of exoplanets and their potential for life. 

\acknowledgments
Datasets for this research are available in these in-text data citation references: \citeA{fu2017,crossfield2017,libbyroberts2020,kreidberg2020,best2020zenodo,parmentier2016,beatty2019,gao2021zenodo}. We are grateful for NASA’s Astrophysics Data System, without which this review would not have been possible. We thank X. Zhang and J. J. Fortney for valuable discussions about the structure of this review. We also thank I. J. M. Crossfield, L. Kreidberg, Y. Kawashima, and C. He for contributing to the production of several figures. P. G. acknowledges support from H. Zhang, W. Z. Gao, H. Y. Dai, H. P. Zhang, the 51 Pegasi b Fellowship sponsored by the Heising-Simons Foundation, and NASA through the NASA Hubble Fellowship grant HST-HF2-51456.001-A awarded by the Space Telescope Science Institute, which is operated by the Association of Universities for Research in Astronomy, Inc., for NASA, under contract NAS5-26555. S. E. M. acknowledges support from NASA Earth and Space Science Fellowship Grant 80NSSC18K1109.

\bibliography{references}

\end{document}